\documentclass[aps,floats,floatfix,showpacs,amssymb,prd,twocolumn,superscriptaddress,nofootinbib,longbibliography,reprint]{revtex4-2}

\usepackage{amssymb,amsmath,verbatim,mathtools,needspace,enumitem,etoolbox,graphicx,physics,microtype,afterpage,xspace,tabularx,lmodern,multirow,mathrsfs}
\usepackage{gensymb}
\usepackage[normalem]{ulem}
\usepackage[dvipsnames, usenames]{xcolor}
\definecolor{linkcolor}{rgb}{0.0,0.3,0.5}
\usepackage[unicode, colorlinks=true, linkcolor=linkcolor, citecolor=linkcolor, filecolor=linkcolor, urlcolor=linkcolor, linktocpage, breaklinks]{hyperref}
\usepackage[all]{hypcap}
\usepackage{subfigure}
\usepackage[T1]{fontenc}
\usepackage[utf8]{inputenc}
\usepackage[usenames,dvipsnames]{xcolor}
\usepackage{upgreek}
\usepackage{booktabs}

\usepackage{braket}
\usepackage[linesnumbered,ruled]{algorithm2e}

\usepackage{orcidlink}

\graphicspath{{figs/}}

\hypersetup{colorlinks=true,citecolor=romared,linkcolor=romared,urlcolor=romared}

\definecolor{romared}{RGB}{142,0,28}

\newcommand{\bea}{\begin{eqnarray}}
\newcommand{\eea}{\end{eqnarray}}
\newcommand{\be}{\begin{equation}}
\newcommand{\ee}{\end{equation}}
\newcommand{\ba}{\begin{align}}
\newcommand{\ea}{\end{align}}
\newcommand{\nn}{\notag \\}

\newcommand{\scri}{\ensuremath{\mathscr{I}}}
\newcommand{\isco}{\text{ISCO}}

\newcommand{\spin}{\frak s}
\newcommand{\rh}{r_{\rm h}}
\newcommand{\rc}{r_{\rm c}}

\newcommand{\calP}{\mathrm{P}}
\newcommand{\calR}{\mathrm{R}}
\newcommand{\e}{\varepsilon}
\newcommand{\That}{{\mathcal{T}}}

\begin{document}

\title{Multi-domain spectral method for Kerr spacetimes: \\ a hyperboloidal Teukolsky $m$-mode solver for self-force computations}

\author{Brad Cownden\, \orcidlink{0000-0002-5831-1384}}
\affiliation{Instituto Galego de F\'{\i}sica de Altas Enerx\'{\i}as, Universidade de Santiago de Compostela, Santiago de Compostela, Spain}

\author{Rodrigo Panosso Macedo}
\affiliation{Center of Gravity, Niels Bohr Institute, Blegdamsvej 17, 2100 Copenhagen, Denmark}

\author{Christiana Pantelidou}
\affiliation{School of Mathematics and Statistics, University College Dublin, Dublin, Ireland}

\author{Samuel D.\ Upton\,\orcidlink{0000-0003-2965-7674}}
\affiliation{School of Mathematics and Statistics, University College Dublin, Dublin, Ireland}
\affiliation{School of Mathematical Sciences and STAG Research Centre,
University of Southampton, Southampton, SO17 1BJ, United Kingdom}

\author{Barry Wardell}
\affiliation{School of Mathematics and Statistics, University College Dublin, Dublin, Ireland}

\begin{abstract}
We solve the frequency-domain, $m$-mode Teukolsky equation for the spin-weight $\spin=-2$ Weyl scalar $\psi_4$ sourced by a point mass on a circular, equatorial orbit around a Kerr black hole. The equation is posed on hyperboloidal slices that reach the event horizon and future null infinity, so that the physical (retarded) solution is selected by regularity alone, without imposing radiation conditions. The particle is treated with the $m$-mode puncture/effective-source scheme, and the resulting two-dimensional elliptic problem is solved with a four-domain pseudo-spectral method in which the equatorial symmetry is broken and handled by an even/odd decomposition of the field. 
We test the method for orbital radius $r_0=10M$ and spins up to $|a|/M=0.9$, covering prograde and retrograde orbits. The energy fluxes to infinity agree with independent Black Hole Perturbation Toolkit results to $\sim10^{-7}$ for the dominant modes and to better than $2\times10^{-5}$ for all modes up to $m=20$ at moderate resolutions. The static mode $m=0$ requires no special treatment.
\end{abstract}

\maketitle
\tableofcontents

\section{Introduction}


Space-based gravitational-wave detectors such as LISA will observe extreme- and intermediate-mass-ratio inspirals (EMRIs and IMRIs), in which a compact object spirals into a black hole that is hundreds to millions of times more massive~\cite{Babak:2017tow,LISA:2022yao}. Such systems complete $10^4$--$10^5$ orbits in the detector band, and extracting their science requires waveform models whose phase is accurate over the whole inspiral. The natural framework for addressing such systems is gravitational self-force theory~\cite{Barack:2018yvs,Pound:2021qin}, in which the field equations are expanded in the small mass ratio $\varepsilon$ and solved order by order. Phase accuracy over the inspiral requires going to second order in $\varepsilon$~\cite{Hinderer:2008dm,Burke:2023lno}. At second order, the binding energy~\cite{Pound:2019lzj}, the energy flux~\cite{Warburton:2021kwk} and complete waveforms~\cite{Wardell:2021fyy} are now available for quasicircular inspirals into a Schwarzschild black hole, including a spinning secondary~\cite{Mathews:2021rod}. Astrophysical black holes, however, spin, and the extension of these calculations to a Kerr background is one of the main open problems of the field. This work represents an important step towards addressing that problem.

Most self-force calculations decompose the fields into spherical or spheroidal harmonic $(\ell,m)$ modes. In the frequency domain this reduces the field equations to relatively simple ordinary differential equations~\cite{Pound:2021qin}. The price paid for this simplification is that the singular nature of the solution leads to a sum over $(\ell,m)$ modes (required to produce the full field) that converges very slowly in vicinity of the worldline of the smaller body. At second order this becomes a major bottleneck: the source of the second-order field equation is quadratic in the first-order field, and the poor convergence means we require an excessively large number of first-order $(\ell,m)$ modes~\cite{Miller:2016hjv}. This has historically been the largest bottleneck in second-order calculations~\cite{Upton:2025bja}.

An alternative is to decompose the fields into azimuthal modes $e^{im\varphi}$ only, and to solve two-dimensional partial differential equations in the radial and polar directions.
Such $m$-mode schemes were introduced for the Lorenz-gauge equations in Kerr, which do not separate in the polar angle, in the time domain~\cite{Barack:2007we,Dolan:2010mt,Dolan:2011dx,Dolan:2012jg,Thornburg:2016msc} and, more recently, in the frequency domain~\cite{Osburn:2022bby,Osburn:2026uct}.
In Ref.~\cite{PanossoMacedo:2024pox}, a subset of us developed a multi-domain spectral solver for the $m$-mode equations, demonstrated for a scalar charge on a circular orbit around a Schwarzschild black hole.
It combines a puncture (effective-source) scheme~\cite{Vega:2007mc,Warburton:2013lea} (calculated in Ref.~\cite{Bourg:2024cgh}), a domain decomposition with a particle domain adapted to the comoving coordinates of the particle, and the hyperboloidal frequency-domain framework of Ref.~\cite{PanossoMacedo:2022fdi}.
In the effective-source scheme, one takes a puncture field that captures the multipole structure of the small body and diverges on the worldline of the object, and uses that to solve for a smoother residual field which determines the motion of the body.

In this paper we extend the approach of Ref.~\cite{PanossoMacedo:2024pox} to the gravitational case in Kerr spacetime. Specifically, we solve the frequency-domain, $m$-mode Teukolsky equation~\cite{Teukolsky:1973ha} for the spin-weight $\spin=-2$ Weyl scalar $\psi_4$, sourced by a point mass on a circular, equatorial orbit. Although the Teukolsky equation separates, its angular functions are frequency-dependent spheroidal harmonics, and an $m$-mode formulation is the one that carries over to the second-order sources~\cite{Spiers:2023cip}. $\psi_4$ also directly provides the gravitational-wave fluxes and, through CCK metric reconstruction~\cite{Chrzanowski:1975wv,Kegeles:1979an}, the metric perturbation.\footnote{While this reconstruction procedure is only valid in vacuum regions of spacetime, GHZ reconstruction~\cite{Green:2019nam,Toomani:2021jlo,Hollands:2024iqp} builds upon this method to allow the construction of metric perturbations with sources via the inclusion of a specific corrector tensor.
A puncture scheme for the Teukolsky equation, within the GHZ approach to metric reconstruction, has been implemented in $\ell m$-modes in Schwarzschild in Ref.~\cite{Bourg:2024vre}.}
In this work the Teukolsky equation is posed on hyperboloidal slices that extend from the future event horizon to future null infinity~\cite{Zenginoglu:2007jw,Zenginoglu:2011jz,Ansorg:2016ztf,PanossoMacedo:2019npm,PanossoMacedo:2023qzp}. On these slices the physical, retarded solution is selected by regularity at the boundaries, without imposing radiation conditions or matching to asymptotic solutions, and the waveform and the horizon flux are read off directly at the boundaries of the numerical domain. A two-dimensional hyperboloidal treatment of the Teukolsky operator has also been used to compute Kerr quasinormal modes~\cite{Assaad:2025nbv}.
In our effective source, we use a puncture for $\psi_4$ that is constructed from the Lorenz gauge puncture presented in Ref.~\cite{lmtompaper}.
While we focus on solving the linearised problem in this paper, we note that many of the techniques we employ here will carry over to second order; see Sec~\ref{sec:conclusion} for further discussion.

The step from a scalar field in Schwarzschild to $\psi_4$ in Kerr is not only a change of operator. Both the spin weight of the field and the spin of the black hole break the reflection symmetry across the equatorial plane. We handle this with a decomposition of the field into parts even and odd under $x=\cos\theta\to-x$, which keeps $y=\cos^2\theta$ as the polar coordinate in which the solution is smooth across the equator. At the symmetry axis, the field equation alone does not select the regular solution of the spin-weighted problem: after the standard regularisation, both local branches are smooth, and the regular one must be imposed through Dirichlet conditions at the poles. Without them, the discrete problem has a near null space and the solution does not converge with resolution.  We also replace the iterative solver of Ref.~\cite{PanossoMacedo:2024pox} by a sparse direct solver acting on a matrix assembled directly from the spectral differentiation matrices, which makes a single $m$-mode solve significantly faster.

We test the method primarily for an orbit at $r_0=10M$ around black holes with $a/M=0,\pm0.5,\pm0.9$, covering prograde and retrograde orbits, and modes up to $m=20$. The energy fluxes to future null infinity agree with independent results from the Black Hole Perturbation Toolkit (BHPToolkit)~\cite{BHPToolkit} to $\sim10^{-7}$ for the dominant modes and to better than $2\times10^{-5}$ for all modes, and we obtain the dissipative components of the first-order self-force from the flux balance, with the sum over $m$ accurate to $\sim10^{-7}$. The static mode $m=0$, which requires special treatment in many $m$-mode and frequency-domain approaches, is solved in the same way as all other modes, with no special treatment. The near-extremal limit is also readily available, with spin $a=0.9999$ proceeding smoothly. The overall accuracy of code is currently limited by the order of the puncture.

The paper is organised as follows. Section~\ref{sec:geometry} reviews the Teukolsky equation in Kerr spacetime, introduces the hyperboloidal coordinates and the regularised field, and analyses its behaviour at the axis, the equator, the horizon and null infinity. Section~\ref{sec:particle} describes the orbit, the comoving coordinates and the puncture and effective source. Section~\ref{sec:mdsm} introduces the even/odd decomposition, the four domains and the boundary and interface conditions, and Sec.~\ref{sec:spectral} the spectral discretisation and the linear solver. Our results, including the comparison with the Black Hole Perturbation Toolkit, the dissipative self-force and timings, are in Sec.~\ref{sec:results}. We conclude in Sec.~\ref{sec:conclusion}.

\subsection*{Notation}
We employ lower case Latin letters to represent tensorial quantities in the abstract index notation, whereas Greek letters indicate the components of the tensor in a given coordinate system. Thus, $u^a$ is a vector, whereas $u^{ \mu}$ or $u^{\bar \mu}$ are its components expressed, for instance, in the Boyer--Lindquist coordinates $x^\mu = (t,r,\theta,\phi)$, or hyperboloidal coordinates $\bar x^\mu = (\tau, \sigma, x,\varphi)$.
When changing coordinates, we will abuse the notation for functions originally defined on $x^\mu$, but then re-parametrised in $\bar x^\mu$, i.e., for a given $f(x^\mu)$,  the notation $f(\bar x^\mu)$ should be understood as the result from $f(x^\mu(\bar x^\nu))$.
With the symbol $\tau$ reserved to the hyperboloidal time coordinate, we employ $\uptau$ for the particle's proper time.
The hyperboloidal coordinates are dimensionless, and we use $\lambda$ to represent a typical spacetime length scale. The spin-weight of a field reads $\spin$, while the letter $s$ is reserved for a dimensionless (Laplace) frequency, related to the usual Fourier frequency via $s = - i \lambda \omega$.

The Newman--Penrose components of the Weyl tensor $C_{abcd}$ in the physical spacetime are denoted by $\psi_n$ ($n=0,\ldots,4$), whereas $\Psi_\spin$ are the Teukolsky master functions, with $\Psi_2=\psi_0$ and $\Psi_{-2}=\zeta^4\psi_4$.
Quantities defined in the conformal spacetime, with metric $\bar g_{ab}=\Omega^2 g_{ab}$, or expressed in the hyperboloidal coordinates $\bar x^\mu$, carry an overline bar.
The Weyl tensor with one contravariant index is conformally invariant, $\bar C^a{}_{bcd}=C^a{}_{bcd}$, hence $\bar C_{abcd}=\Omega^2 C_{abcd}$. The rescaled conformal Weyl tensor is $\bar d^a{}_{bcd}=\Omega^{-1}\bar C^a{}_{bcd}$, and its Newman--Penrose components with respect to a conformal null tetrad are denoted by $\bar\phi_n$ ($n=0,\ldots,4$).
The field solved for numerically follows from $\Psi_\spin$ through a sequence of steps, summarised in Table~\ref{tab:notation}: the conformal rescaling to $\bar\Psi_\spin$ (Sec.~\ref{sec:Hyp_Teuk}), the frequency-domain $m$-mode decomposition $\bar\Phi_{\spin m}$, the regularisation at the axis $\bar\psi_{\spin m}$ (Sec.~\ref{sec:reg_ang}), the split into residual and puncture fields near the particle (Sec.~\ref{sec:puncture}), and the even/odd decomposition (Sec.~\ref{sec:mdsm}). Each step carries its own operator and source term, and all numerical results are expressed in terms of the axis-regularised field $\bar\psi_{\spin m}$, its residual part, or its even/odd components.

\begin{table}[t]
\centering
\footnotesize
\setlength{\tabcolsep}{6pt}
\renewcommand{\arraystretch}{1.25}
\begin{tabular}{llll}
\hline\hline
Quantity & Field & Operator & Source \\
\hline
Weyl scalars & $\psi_n$ & -- & -- \\
Master function & $\Psi_\spin$ & $\mathcal O_\spin$ & $S_\spin$ \\
Rescaled conf.\ Weyl scalars & $\bar\phi_n$ & -- & -- \\
Conformal master function & $\bar\Psi_\spin$ & $\bar{\mathcal O}_\spin$ & $\bar S_\spin$ \\
Conformal $m$-mode & $\bar\Phi_{\spin m}$ & $\bar{\cal D}_{\spin ms}$ & $\bar{\rm S}_{\spin m}$ \\
Axis-regularised & $\bar\psi_{\spin m}$ & $\bar{\frak D}_{\spin ms}$ & $\bar{\frak S}_{\spin m}$ \\
Residual / puncture & $\bar\psi^{\rm R}_{\spin m}$, $\bar\psi^{\rm P}_{\spin m}$ & $\bar{\frak D}_{\spin ms}$ & $\bar{\frak S}^{\rm eff}_{\spin m}$ \\
Even/odd & $\bar\psi^{\rm E/O}_{\spin m}$ & $\hat{\bar{\frak D}}^{\rm E/O}_{\spin ms}$ & $\bar{\frak S}^{\rm E/O}_{\spin m}$ \\
\hline\hline
\end{tabular}
\caption{Notation for the fields, operators and source terms, from the physical Weyl scalars to the variables used in the numerical solver. Quantities without an overline refer to the physical spacetime, and barred quantities to the conformal spacetime in hyperboloidal coordinates. From Sec.~\ref{sec:mdsm} onwards, the indices $\spin m$ are omitted.}
\label{tab:notation}
\end{table}



\section{Teukolsky equation in hyperboloidal coordinates}\label{sec:geometry}

We begin by expanding the full metric around the background Kerr spacetime in powers of the mass ratio $\e$, so that
\begin{equation}
    g^{\text{exact}}_{ab} = g_{ab} + \e h^{(1)}_{ab} + \mathcal{O}(\e^2).
\end{equation}
The first-order metric perturbation then satisfies the linearised Einstein field equations sourced by a point mass:
\begin{equation}
    G^{(1)}_{ab}[h^{(1)}] = 8\pi T_{ab}.
\end{equation}
By acting on the metric perturbation with the linear operator $\That$~\cite{Pound:2021qin}, we can construct the perturbation to the Weyl scalar
\begin{equation}
    \psi_{4} = \That_{4}[h^{(1)}], \label{eq:Th1}
\end{equation}
which satisfies the Teukolsky equation
\begin{equation}
    \mathcal{O}_{-2}[\zeta^4\psi_{4}] = S_{-2},
\end{equation}
where $\mathcal{O}_{\spin}$ is the spin-weight $\spin$ Teukolsky operator given below in Eq.~\eqref{eq:Teuk_Op}, $\zeta$ is the Killing--Yano coefficient and $S_{-2}$ is the source constructed from derivatives of the stress-energy tensor.



\subsection{Kerr coordinates}

Consider a Kerr black hole with mass $M$ and angular momentum $Ma$ with spacetime metric
\begin{align}
    \label{e: Kerr BL}
    ds^2 &= -f dt^2 - \dfrac{4M ar}{\Sigma} \sin^2 \theta \, dt d\phi + \dfrac{\Sigma}{\Delta} \, dr^2 + \Sigma \, d\theta^2 \nonumber \\
    &\qquad + \sin^2 \theta \left(r^2 + a^2 + \dfrac{2M a^2 r}{\Sigma} \sin^2 \theta \right) d\phi^2
\end{align}
where $x^\mu = (t, r, \theta, \phi)$ are the Boyer--Lindquist coordinates, with the metric functions
\begin{align}
    f(r,\theta) &= 1 - \dfrac{2M r}{\Sigma(r,\theta)}, \quad     \Sigma(r, \theta) = r^2 + a^2 \cos^2 \theta \, , \nn
    \Delta(r) &= r^2 - 2M r + a^2 = (r - \rh)(r - \rc).
\end{align}
The event and Cauchy horizons are, respectively,  given by
\bea
\rh &=& M + \sqrt{M^2 - a^2}, \\
\rc &=& M - \sqrt{M^2 - a^2}.
\eea
We define the tortoise coordinate $r_*$ and phase $\phi_*$ as
\begin{align}
\label{eq:tortoises}
    \dfrac{d r_*}{d r} = \dfrac{r^2 + a^2}{\Delta} \quad \text{and} \quad \dfrac{d \phi_*}{d r} = \dfrac{a}{\Delta} \, .
\end{align}
We choose to parametrize dimensionful quantities in terms of black hole horizon length $\rh$ as opposed to the black hole mass $M$. Thus, we introduce the dimensionless rotation parameter, equivalent to the ratio between the Cauchy and event horizons via
\be
\kappa = \dfrac{a}{\rh} \Longleftrightarrow  \kappa^2 = \dfrac{\rc}{\rh}.
\ee
In terms of $\rh$, the mass is parameterised by
\begin{align}
   M = \dfrac{r_h}{2}(1+ \kappa^2).
\end{align}
Additionally the horizons' angular velocities read
\begin{align}
    \Omega_{\rm h} ={}& \dfrac{a}{2M \rh} = \dfrac{\kappa}{\rh (1+\kappa^2)}, \\
    \Omega_{\rm c} ={}& \dfrac{a}{2M \rc} = \dfrac{1}{\rh \kappa (1+\kappa^2)},
\end{align}
and the horizons' surface gravity is
\begin{align}
    \varkappa_{\rm h} ={}& \dfrac{1}{4M} - M \Omega_{\rm h} ^2 = \dfrac{1-\kappa^2}{2\rh (1+\kappa^2)} , \\
    \varkappa_{\rm c} ={}& \dfrac{1}{4M} - M \Omega_{\rm c} ^2 = - \dfrac{1-\kappa^2}{2\rh \kappa^2 (1+\kappa^2)}.
\end{align}
In terms of the quantities above, Eq.~\eqref{eq:tortoises} integrates to
\begin{align}
    r_*(r) ={}& r + 2 M \ln\left( \dfrac{r}{\rh} \right) \nonumber \\
        & + \dfrac{1}{2 \varkappa_{\rm h}} \ln\left(1-  \dfrac{\rh}{r} \right)  + \dfrac{1}{2 \varkappa_{\rm c}} \ln\left(1 -  \dfrac{\rc}{r} \right), \label{eq:tort_r} \\
    \phi_*(r) ={}& \dfrac{\Omega_{\rm h}}{2\varkappa_{\rm h}} \ln\left(1-  \dfrac{\rh}{r} \right) + \dfrac{\Omega_{\rm c}}{2\varkappa_{\rm c}} \ln\left(1-  \dfrac{\rc}{r} \right). \label{eq:tort_chi}
\end{align}
\subsubsection{Teukolsky Equation}
In these coordinates, the Teukolsky equation takes the form $\mathcal{O}_\spin \Psi_\spin =S_\spin$ with the spin-weighted Teukolsky operator
\begin{align}
\label{eq:Teuk_Op}
    \mathcal{O}_\spin &:=-\left(\dfrac{(r^2+a^2)^2}{\Delta} - a^2 (1-x^2)\right)\partial_t^2 - \dfrac{4 M a r}{\Delta} \partial_t \partial_\phi \nn
    &- \left(\dfrac{a^2}{\Delta} - \dfrac{1}{1-x^2}\right)\partial_\phi^2 + \Delta \partial_r^2 + 2(1+\spin)(r-M)\partial_r \nn
    & + (1-x^2)\partial_x^2 - 2x\partial_x + 2\spin\left(\dfrac{M(r^2-a^2)}{\Delta} - r-iax \right)\partial_t \nn
    &
    +2 \spin\left(\dfrac{a(r-M)}{\Delta} + \dfrac{ix}{1-x^2} \right)\partial_\phi
    - \spin\left(\dfrac{\spin x^2}{1-x^2} - 1\right),
\end{align}
where the angular coordinate was transformed to $x=\cos\theta$. The radial first-derivative term is $\Delta^{-\spin}\partial_r(\Delta^{\spin+1}\partial_r) - \Delta\partial_r^2 = (1+\spin)\Delta'(r)\partial_r$.  The source term takes the form $S_\spin=-4\pi \Sigma T_\spin$, where $T_\spin$ is the stress-energy for the point-particle representation of the secondary black hole, cf.~Sec.~\ref{sec:particle}.

We recall that the master functions $\Psi_\spin$'s are directly related to the Weyl scalars $\psi_0$ and $\psi_4$, via
\be
\Psi_2 = \psi_0, \quad \Psi_{-2} = \zeta^4 \psi_4,
\ee
with $\zeta(r,\theta) = r - i a \cos\theta$ the Killing--Yano coefficient in Boyer--Lindquist coordinates.
Specifically, $\psi_0$ and $\psi_4$ are the components of the Weyl tensor $C_{abcd}$ projected into the Kinnersley tetrad basis
\begin{align}
\ell^\mu ={}& \dfrac{r^2+a^2}{\Delta(r)}\delta^\mu_t + \delta^\mu_r + \dfrac{a}{\Delta(r)}\delta^\mu_\phi, \label{eq:Kin_null_tetrad_lk} \\
k^\mu ={}&\dfrac{1}{2\Sigma(r, \theta)}\bigl[ \left(r^2+a^2\right)\delta^\mu_t -\Delta(r) \, \delta^\mu_r + a\,\delta^\mu_\phi\bigr], \\
m^\mu ={}& \dfrac{1}{{\sqrt{2} (r+ i a \cos\theta)}} \left( i a \sin\theta \delta^\mu_t + \delta^\mu_\theta + \dfrac{i}{\sin\theta}\delta^\mu_\phi \right). \label{eq:Kin_null_tetrad_m}
\end{align}
according to the Newman--Penrose formalism~\cite{NewmanPenrose1962}.

We are interested in the retarded solution to the Teukolsky equation, i.e., solutions $\Psi^{\rm ret}_\spin$ satisfying boundary conditions with purely outgoing modes in the wave zone as $r\to \infty$ ($r_*\to \infty$), and purely ingoing modes at the horizon as $r\to \rh$ ($r_*\to -\infty$), see e.g.~Ref.~\cite{Pound:2021qin} for more details. We impose this behaviour geometrically via the hyperboloidal framework~\cite{Zenginoglu:2011jz,PanossoMacedo:2019npm,PanossoMacedo:2023qzp}, and we refrain from using the superscript ${}^{\rm ret}$, unless necessary for comprehension.

\subsection{Hyperboloidal Coordinates}
The hyperboloidal framework provides a coordinate system well adapted to account for the asymptotic behaviour at $r_* \to \pm\infty$~\cite{Zenginoglu:2007jw,Zenginoglu:2011jz}. Employing the notation from Refs.~\cite{PanossoMacedo:2023qzp,Minucci:2024qrn,Assaad:2025nbv}, a hyperboloidal foliation $\bar x^\mu = (\tau, \sigma, x, \varphi)$ follows from the Boyer--Lindquist  $x^\mu = (t,r,\theta,\phi)$ via~\cite{Zenginoglu:2007jw, Zenginoglu:2011jz}
\begin{align}
	\label{e: hyp}
    t ={}& \lambda \big(\tau - H(\sigma) \big) \, , & r ={}& r_o + \dfrac{r_h - r_o}{\sigma} , \\
    \label{eq:hyp_ang}
    x={}&\cos(\theta), & \phi ={}& \varphi - \varphi_*(\sigma),
\end{align}
where $\lambda$ is a given length scale of the spacetime. This parameter is suitable to adapt the formalism across different normalizations in the literature. Setting $M=1$ is equivalent to choosing $\lambda=M$. Here, we will eventually employ $\lambda = \rh$, whereas some works such as Ref.~\cite{Ansorg:2016ztf, PanossoMacedo:2022fdi,Bourg:2025lpd} use the surface gravity $\lambda = \varkappa_{\rm h}^{-1} (= 4M$ in the Schwarzschild case).

A hyperboloidal foliation arises when the height function $H(\sigma)$ in Eq.~\eqref{e: hyp}
satisfies specific asymptotic conditions~\cite{PanossoMacedo:2019npm}. The choice
\begin{align}
    H(\sigma) ={}& {-}\dfrac{\rh - r_o}{\lambda\,\sigma} + \dfrac{2M}{\lambda} \ln \sigma \nn
        & + \dfrac{1}{2 \lambda \varkappa_h } \ln(1- \sigma) + \dfrac{1}{2 \lambda \varkappa_c } \ln\left(1- \dfrac{\sigma}{\sigma_c}\right), \label{eq:Height_MinGauge}
\end{align}
retains the minimal structure ensuring {$\tau=$ constant} to be a hyperboloidal hypersurface.

The radial transformation in Eq.~\eqref{e: hyp} includes only the minimal structure
for a radial compactification $r\sim 1/\sigma$. Apart from mapping $r\rightarrow \infty$ to $\sigma=0$ and $r=\rh$ to $\sigma =1$, it also allows for a generic offset $r_o$. This parameter permits us to control the location of the Cauchy horizon
\be
\sigma_{\rm c} =\dfrac{\rh - r_o}{\rc-r_o} \Longleftrightarrow \rho_o := \dfrac{r_o}{\rh} = \dfrac{1-\kappa^2 \sigma_{\rm c}}{1- \sigma_{\rm c}}.
\ee
Two choices are of relevance~\cite{PanossoMacedo:2019npm,Assaad:2025nbv}:
\begin{enumerate}
    \item {\bf Radial Fixing:}

    $r_o=0$ ($\rho_o = 0$) yields $\sigma_{\rm c} = \rh/\rc =\kappa^{-2}\geq 1$.
    \item {\bf Cauchy Fixing:}

    $r_o =\rc$ ($\rho_o = \kappa^2$) fixes $\sigma_{\rm c} \to \infty$, regardless of $\kappa$.
\end{enumerate}
These options lead to different extremal limits \cite{PanossoMacedo:2019npm,Minucci:2024qrn,Assaad:2025nbv}. 

All results in this paper use radial fixing, $\rho_o = 0$, together with $\lambda = \rh$, so that $r = \rh/\sigma$. Two reasons motivate this choice. First, radial fixing has a continuous limit as $|a|\to M$, so neither the hyperboloidal geometry nor the upcoming elliptic system restricts the treatment of extremal black holes. Second, for the homogeneous problem, the Kerr quasinormal mode spectra and eigenfunctions obtained from the two-dimensional hyperboloidal Teukolsky operator show the same accuracy and convergence in both gauges~\cite{Assaad:2025nbv}, and we expect the same to hold for the sourced problem. The Cauchy fixing yields more compact coefficients, however carries divergences of $(1-\kappa^2)^{-1}$. The expressions that follow, in particular in Sec.~\ref{sec:Hyp_Teuk}, keep $r_o$ generic, so that this alternative can be explored in future works.

In the angular sector, Eq.~\eqref{eq:hyp_ang} explicitly introduces the mapping for the polar angle $\theta$ already used in the Teukolsky Eq.~\eqref{eq:Teuk_Op}. As in the case of ingoing Kerr coordinates, Eq.~\eqref{eq:hyp_ang} also introduces the well-known transformation in the azimuthal angle with
\begin{align}
    \varphi_*(\sigma) ={}& \phi_*(r(\sigma)) \nn
        ={}& \dfrac{\Omega_h}{2 \varkappa_h } \ln(1- \sigma) + \dfrac{\Omega_c}{2  \varkappa_c } \ln\left(1- \dfrac{\sigma}{\sigma_c}\right),
\end{align}
necessary for regularity across the BH horizon.

Implementing the hyperboloidal transformation \eqref{e: hyp} allows for a straightforward conformal re-scaling of the Kerr spacetime~\cite{PanossoMacedo:2019npm, Minucci:2024qrn, Assaad:2025nbv}
\begin{equation}
\label{eq:metric_conformal_trasfo}
    ds^2 = \Omega^{-2} d\bar s^2\,,\quad \Omega(\sigma) = \dfrac{\sigma}{\lambda}.
\end{equation}
Introducing~\cite{PanossoMacedo:2019npm}
\be
\label{eq:conf_KerrFunc}
\bar \Sigma(\sigma, x) = \Omega^2(\sigma) \, \Sigma(\sigma,x), \quad \bar \Delta(\sigma) = \Omega^2(\sigma) \, \Delta(\sigma),
\ee
the (inverse) conformal metric is explicitly given by
\be
\label{eq:conf_metric_hyp}
\bar g^{\bar \mu \bar \nu} = \dfrac{1}{\bar \Sigma(\sigma, x)} \left(
\begin{array}{cccc}
    -W(\sigma,x) & \gamma_\sigma(\sigma) &  0 &  \alpha(\sigma) \\
    {\rm sym.} & p(\sigma) & 0  & \gamma_\varphi(\sigma)\\
    {\rm sym.} & {\rm sym.} &1-x^2 & 0\\
    {\rm sym.} & {\rm sym.} & {\rm sym.} & (1-x^2)^{-1}
\end{array}
\right).
\ee
with metric components
\begin{align}
    p(\sigma) ={}&   \sigma^2(1-\sigma)\left(1-\sigma/\sigma_{\rm c}\right), \label{eq:p_def} \\
    \alpha(\sigma) ={}& {-}\kappa \dfrac{r_h }{\lambda(1- \rho_o)}
\bigl[ 1 + 2\sigma \left( 1+ \kappa^2 \right)  -\rho_o \bigr], \label{eq:alpha_def} \\
    \gamma_\varphi(\sigma) ={}& {-}\kappa \, \dfrac{\sigma^2}{1-\rho_o}, \label{eq:gamma_varphi_def} \\
    \gamma_\sigma(\sigma) ={}& \dfrac{r_h}{\lambda(1-\rho_o)}\bigl\{ 1 - 2\sigma^2 + (1-\sigma)^2 \rho_o^2 \nn
        & - \kappa^2\sigma^2\bigl[1 + 2(1-\sigma)(1+\kappa^2)\bigr] \nn
        & - 2(1-\sigma)\bigl[1 - \sigma^2 (1 + \kappa^2) \bigr]\rho_o\bigr\}, \label{eq:gamma_sigma_def} \\
    W(\sigma, x) ={}& w(\sigma) + \varpi(x), \label{eq:W_def}
\end{align}
where
\begin{align}
    \varpi(x) ={}& {-} \kappa^2\,\dfrac{\rh^2}{\lambda^2}\,(1-x^2), \\
    w(\sigma) ={}& \dfrac{4 r_h^2}{\lambda^2}\, (1+\kappa^2)\, \bigg\{ (1+\sigma)(1+\kappa^2) - \kappa^2 \sigma^2 \nonumber \\
    & - \rho_o\biggl[ 1-\sigma -\dfrac{\sigma(\sigma+\kappa^2)}{1-\rho_o}\biggr] + \dfrac{ \sigma (1-\sigma)\kappa^4}{1-\rho_o} \bigg\}.
\end{align}
One particular aspect to note is the choice to keep $\sim \sin^2\theta = 1-x^2$ in the angular functions $\varpi(x)$, as opposed to just retaining $\sim\cos^2\theta=x^2$ in $\varpi(x)$ and shifting the constant $-\kappa^2 \rh^2/\lambda^2$ to $w(\sigma)$. This option impacts the separation of variables in the frequency domain, yielding an angular contribution to the differential operator slightly different from the ones adopted in the reviews~\cite{Pound:2021qin, Berti:2025hly}.

Expressing the metric functions in terms of the inverse metric has the advantage that they directly relate to most terms in the hyperboloidal Teukolsky equation, as discussed in the next section.

\subsubsection{Teukolsky operator}\label{sec:Hyp_Teuk}
The hyperboloidal Teukolsky master function $\bar \Psi_\spin$ relates to the original $\Psi_\spin$ via~\cite{Zenginoglu:2011jz,PanossoMacedo:2019npm}
\be
\label{eq:hyp_Teuk}
\Psi_\spin(x^\mu) = \lambda^\spin\, \Omega^{1+2\spin} \bar\Delta^{-\spin}\, \bar \Psi_\spin(\bar x^\mu),
\ee
with $\bar \Delta$ defined in Eq.~\eqref{eq:conf_KerrFunc}. In other words, apart from the coordinate change from Boyer--Lindquist $x^\mu$ into the hyperboloidal coordinates $\bar x^\mu$, the master function $\bar \Psi_\spin(\bar x^\mu)$ also re-scales with terms necessary to remove explicit singular terms in the resulting hyperboloidal Teukolsky equation~\cite{Zenginoglu:2011jz,PanossoMacedo:2019npm}.

Geometrically, these pre-factors naturally arise when boosting the Kinnersley null tetrad to ensure regularity across the black hole horizon, and identifying $\bar \Psi_\spin$ with the re-scaled conformal Weyl scalar~\cite{Minucci:2024qrn, Gasperin:2026znw}.

Specifically, the re-scaled conformal Weyl tensor is defined as $\bar d_{abcd} = \Omega C_{abcd}$. Then, equivalently to $\psi_0$ and $\psi_4$, the re-scaled conformal Weyl scalars $\bar\phi_0$ and $\bar\phi_4$ are the Newman--Penrose components of $\bar d_{abcd}$ upon projection into a given conformal null tetrad basis~\cite{Gasperin:2026znw}. The particular conformal hyperboloidal null tetrad employed is related to the original Kinnersley basis~\cite{Minucci:2024qrn} via
\be
\bar \ell^a = \Omega^{-2} \dfrac{\bar \Delta}{\lambda} \, \ell^a, \quad \bar k^a = \dfrac{\lambda}{\bar \Delta} \,k^a, \quad \bar m^a = \Omega^{-1} m^a,
\ee
 Thus, $\bar\phi_0$ and $\bar\phi_4$ assume finite values at $\sigma=0$ (and $\sigma=1$) and the conformal re-scaling recovers exactly the peeling properties for $\psi_0$ and $\psi_4$ at $\scri^+$~\cite{Gasperin:2026znw}. Indeed, Eq.~\eqref{eq:hyp_Teuk} gives $\Psi_2\propto\Omega^5$ and $\Psi_{-2}\propto\Omega^{-3}$ as $\Omega\to0$ towards future null infinity, which implies $\psi_0\propto\Omega^5$ and $\psi_4\propto\Omega$. The $\lambda$ factor ensures all conformal quantities are dimensionless.

In this way, the spin-weighted $\spin=+2$ and $\spin=-2$ conformal master functions relate to the re-scaled conformal Weyl scalars $\bar\phi_0$ and $\bar\phi_4$ via
\be
\bar \Psi_2 = \bar\phi_0, \quad \bar \Psi_{-2} = \bar \zeta^4 \bar\phi_4,
\ee
with
\begin{align}
    \bar \zeta(\sigma,x) ={}& \Omega(\sigma) \zeta(\sigma, x) \\
        ={}& \dfrac{\rh}{\lambda} \bigg( 1 - i \kappa \sigma x + \rho_o(1-\sigma) \bigg)
\end{align}
the conformal Killing--Yano coefficient.

With the above considerations, the conformal Teukolsky equation becomes
$\bar{\mathcal{O}}_\spin \, \bar \Psi_\spin = \bar S_\spin$. As in Eq.~\eqref{eq:hyp_Teuk} the conformal source term relates to the original one via the re-scaling $\bar S_\spin = \lambda^{-\spin}\, \Omega^{-(1+2\spin)} \bar\Delta^{\spin}\, \, S_\spin$.

The conformal Teukolsky operator splits into a part depending only on $\sigma$ and a part depending only on $x$,
\begin{equation}
    \bar{\mathcal{O}}_\spin = \bar{\mathcal{O}}^{(\sigma)}_\spin + \bar{\mathcal{O}}^{(x)}_\spin, \label{eq:hyp_Teuk_time_domain}
\end{equation}
where
\begin{equation}
    \bar{\mathcal{O}}^{(\sigma)}_\spin = \sum_{\alpha\in\mathcal{I}_\sigma} c_\alpha(\sigma)\,\partial_\alpha, \quad
\bar{\mathcal{O}}^{(x)}_\spin = \sum_{\alpha\in\mathcal{I}_x} {\frak c}_\alpha(x)\,\partial_\alpha, 
\end{equation}
with $\partial_{ab}\coloneqq\partial^2_{ab}$, $\partial_0\coloneqq1$, and the index sets
\begin{align}
    \mathcal{I}_\sigma ={}& \{\tau\tau,\sigma\sigma,\tau\sigma,\tau\varphi,\sigma\varphi,\tau,\sigma,\varphi,0\}, \nn
    \mathcal{I}_x ={}& \{\tau\tau,xx,\varphi\varphi,\tau,x,\varphi,0\}.
\end{align}
Thus, $\bar{\mathcal{O}}^{(\sigma)}_\spin$ acts only along $(\tau,\sigma,\varphi)$, and $\bar{\mathcal{O}}^{(x)}_\spin$ only along $(\tau,x,\varphi)$.
The functions depending solely on $\sigma$ read
\begin{subequations}
\begin{align}
    c_{\tau \tau}(\sigma) ={}& {-} w(\sigma), & c_{\sigma \sigma}(\sigma) ={}& p(\sigma), & c_{\tau \sigma}(\sigma) ={}& 2\gamma_\sigma(\sigma), \\
    c_{\tau \varphi}(\sigma) ={}& 2\alpha(\sigma), & c_{\sigma \varphi}(\sigma) ={}& 2\gamma_\varphi(\sigma), & c_\varphi(\sigma) ={}& \gamma_\varphi'(\sigma), \\
    c_{\sigma}(\sigma) ={}& \mathrlap{p'(\sigma) + \spin \,\sigma \big(2 - \sigma\Lambda\big),} \\
    c_\tau(\sigma) ={}& \mathrlap{\gamma_\sigma'(\sigma) + \dfrac{2 \rh \spin}{\lambda}(1+\kappa^2)\big(1 -\sigma\Lambda\big),} \\
    c_0(\sigma) ={}& \mathrlap{{-}\sigma \big[1 +  (1-2\sigma)(\Lambda-1)\big] - \spin\sigma\Lambda,}
\end{align}
\end{subequations}
with
\bea
\Lambda &:=& 1+ \dfrac{1}{\sigma_{\rm c}} \nonumber \\
&=& \left\{
\begin{array}{cc}
1+\kappa^2 = 2M/\rh & \text{(radial fixing)} \\
1 & \text{(Cauchy fixing),} \\
\end{array}\right.
\label{eq:Lambda_def}
\eea
whereas those depending only on $x$ are
\begin{subequations}\label{eq:ang_coeff_eq}
\begin{align}
    {\frak c}_{\tau \tau}(x) ={}& {-} \varpi(x), & {\frak c}_\tau(x) ={}& {-}\dfrac{2\rh}{\lambda} i \kappa x \spin, \\
    {\frak c}_{x x}(x) ={}& 1-x^2, & {\frak c}_{x}(x) ={}& {-}2 x, \\
    {\frak c}_{\varphi \varphi}(x) ={}& \dfrac{1}{1-x^2}, & {\frak c}_\varphi(x) ={}& \dfrac{2 i \spin \, x}{1-x^2}, \\
    {\frak c}_0(x) ={}& \spin - \spin^2\dfrac{x^2}{1-x^2}. &&
\end{align}
\end{subequations}
Equations~\eqref{eq:hyp_Teuk_time_domain}--\eqref{eq:ang_coeff_eq} follow exactly from Eq.~\eqref{eq:Teuk_Op} through the transformations~\eqref{e: hyp}, \eqref{eq:hyp_ang} and the re-scaling~\eqref{eq:hyp_Teuk}, i.e.\ $\bar{\mathcal O}_\spin = \lambda^{-\spin}\Omega^{-(1+2\spin)}\bar\Delta^{\spin}\,\mathcal O_\spin\,\lambda^{\spin}\Omega^{1+2\spin}\bar\Delta^{-\spin}$.

We are now in position to exploit the Killing vectors $\partial_\tau$ and $\partial_\varphi$ to introduce a frequency domain, $m$-mode decomposition via
\begin{align}
    \bar \Psi_{\spin}(\tau, \sigma, x, \varphi) ={}& e^{s \tau + i m \varphi} \, \bar \Phi_{\spin m }(\sigma, x; s), \\
    \bar S_{\spin}(\tau, \sigma, x, \varphi) ={}& e^{s \tau + i m \varphi} \, \bar {\rm S}_{\spin  m }(\sigma, x; s).
\end{align}
For a circular orbit with angular velocity $\Omega_\phi$, each $m$-mode oscillates with $\omega_m = m\,\Omega_\phi$, i.e.\ $s = -i\lambda m\Omega_\phi$. We normalise the $m$-modes as $\Psi_{\spin m}(t,r,\theta)=\int_0^{2\pi}\Psi_\spin\, e^{-im\phi}\,d\phi$, so that $\Psi_\spin = (2\pi)^{-1}\sum_m \Psi_{\spin m} e^{im\phi}$.
Hence, we reduce the problem to the elliptic equation $\bar {\cal D}_{\spin m s}\, \bar \Phi_{\spin  m} = \bar {\rm S}_{\spin m}$. As expected, the $m$-mode frequency domain operator separates into purely radial and purely angular differential operators, i.e.
\begin{align}
\bar {\cal D}_{\spin m s} ={}& \bar{\cal R}_{\spin m s} + \bar{\Xi}_{\spin m s}, \label{eq:Hyp-m-Teukolksy}\\
\bar{\cal R}_{\spin m s} ={}& {\rm R}_{\sigma\sigma}(\sigma)\partial^2_{\sigma \sigma} + {\rm R}_{\sigma}(\sigma)\partial_{\sigma} + {\rm R}_{0}(\sigma), \label{eq:Radial_Operator}\\
\bar{\Xi}_{\spin m s} ={}& {\Gamma}_{xx}(x)\partial^2_{x x} + { \Gamma}_{x}(x)\partial_{x} + {\Gamma}_{0}(x).
\end{align}
The corresponding functions associated with each term follow directly from the original conformal Teukolsky equation \eqref{eq:hyp_Teuk_time_domain}. They read
\begin{align}
    {\rm R}_{\sigma\sigma}(\sigma) ={}& c_{\sigma \sigma}(\sigma), \label{eq:radial_op_functions} \\
    {\rm R}_{\sigma}(\sigma) ={}& c_{\sigma}(\sigma) + s \, c_{\tau\sigma}(\sigma) + i m \, c_{\sigma\varphi}(\sigma), \\
    {\rm R}_0(\sigma) ={}& c_{0}(\sigma) + s \, c_{\tau}(\sigma) + i m \, c_{\varphi}(\sigma) \nn
        & + s^2 c_{\tau\tau}(\sigma) + i\,s\, m\, c_{\tau \varphi}(\sigma),
\end{align}
and
\begin{align}
    {\Gamma}_{xx}(x) ={}& {\frak c}_{x x}(x), \\
    {\Gamma}_{x}(x) ={}& {\frak c}_{x}(x) \\
    {\Gamma}_0(x) ={}& {\frak c}_{0}(x) + s \, {\frak c}_{\tau}(x) + i m \, {\frak c}_{\varphi}(x) + s^2 {\frak c}_{\tau \tau}(x) \nn
        & - m^2 {\frak c}_{\varphi \varphi}(x).
\end{align}
As anticipated in the previous section, the angular differential operator $\bar{\Xi}_{\spin m s}$ differs from the common definition in the literature~\cite{Pound:2021qin, Berti:2025hly}. As a consequence, upon a full separation of variables
\be
\label{eq:func_sep_variables}
\bar \Phi_{\spin  m }(\sigma, x; s) = \bar \Phi_{\spin  \ell m }(\sigma; s) P_{\spin  \ell m }(x; s),
\ee
the function $P_{\spin  \ell m }(x; s)$ would still correspond to the spin-weighted spheroidal harmonic, satisfying
\be
\bar{\Xi}_{\spin m s}  P_{\spin  \ell m } = - \bar A_{\spin  \ell m }(s) P_{\spin  \ell m },
\ee
but the separation constant $\bar A_{\spin  \ell m }(s)$ would differ from the standard $A_{\spin  \ell m }(s)$~\cite{Pound:2021qin, Berti:2025hly} by\footnote{Recall that $a\omega = i \rh\, \kappa\, s /\lambda$}
$
\bar A_{\spin  \ell m }(s) = A_{\spin  \ell m }(s) - (a \omega)^2.
$

Since in this work we do not introduce the full separation of variables~\eqref{eq:func_sep_variables}, the difference in the angular operator and separation constant play no role. However, an analytical knowledge of some properties related to the spin-weighted spheroidal harmonics offers insights about the solution $\bar \Phi_{\spin  m }(\sigma, x; s)$'s behavior at the symmetry axis $x\to \pm 1$ and equator $x=0$.

\subsection{Regularity along the angular direction}\label{sec:reg_ang}

\subsubsection{The symmetry axis \texorpdfstring{$x=\pm 1$}{x=±1}}\label{sec:axis}
The frequency domain hyperboloidal operator still contains a highly singular behaviour at the axis $x=\pm 1$, cf.~Eq.~\eqref{eq:ang_coeff_eq}. A regularization in the angular direction follows through~\cite{PanossoMacedo:2019npm}
\be
\label{eq:reg_axis_ansatz}
\bar \Phi_{\spin  m }(\sigma, x; s) = \bar \psi_{\spin  m }(\sigma, x; s) (1+x)^{-\delta_1/2}(1-x)^{-\delta_2/2},
\ee
with $\delta_1 = |m-\spin|$ and $\delta_2 = |m+\spin|$.

The new field satisfies
\be
\label{eq:aux_field}
\bar {\frak D}_{\spin m s}\,\big[ \bar \psi_{\spin  m}(\sigma, x;s)\big] = \bar {\frak S}_{\spin m}(\sigma, x;s),
\ee
with the re-scaled source
\be
\bar {\frak S}_{\spin m}(\sigma, x; s) = (1+x)^{\delta_1/2}(1-x)^{\delta_2/2}\, \bar {\rm S}_{\spin m s}(\sigma, x; s).
\ee
As before, the elliptic operator $\bar {\frak D}_{\spin m s}$ explicitly splits into radial and angular parts $\bar {\frak D}_{\spin m s} =  \bar{\cal R}_{\spin m s} + \bar{\frak X}_{\spin m s}$ with
\bea
\bar{\frak X}_{\spin m s} = {\frak g}_{xx}(x)\partial^2_{xx} + {\frak g}_{x}(x)\partial_{x} + {\frak g}_{0}(x),
\eea
where the coefficients read
\begin{align}
{\frak g}_{xx}(x) ={}& \Gamma_{xx}(x) = 1-x^2, \label{eq:ang_gxx}\\
{\frak g}_{x}(x) ={}& \Gamma_{x}(x) - \delta_1(1-x) + \delta_2(1+x) \nn
                 ={}& {-}2x - \delta_1(1-x) + \delta_2(1+x),  \label{eq:ang_gx} \\
{\frak g}_{0}(x) ={}& \Gamma_{0}(x) + \dfrac{1-x}{1+x}\dfrac{\delta_1^2}{4} + \dfrac{1+x}{1-x}\dfrac{\delta_2^2}{4} - \dfrac{\delta_1 \delta_2}{2} \nn
                & +\dfrac{\delta_1 +\delta_2}{2} \nn
                ={}&
                \left(\dfrac{\rh}{\lambda} s \kappa\right)^2(1-x^2) - 2 i
                \dfrac{\rh}{\lambda} s \kappa \, \spin x \nn
                & - \dfrac{m^2 -\delta_1-\delta_2 + \delta_1\delta_2 - \spin(2+\spin)}{2}. \label{eq:ang_g0}
\end{align}
Eq.~\eqref{eq:ang_g0} explicitly shows that the terms $\propto \delta_i^2$ combine to cancel out exactly the singular terms from $\Gamma_0(x)\propto (m + x \spin)^2/(1-x^2)$, leaving a regular ${\frak g}_{0}(x)$ at $x\to \pm 1$.

At this point, the separation of variables \eqref{eq:func_sep_variables} offers a support. At $x\to \pm 1$, the spin-weighted spheroidal harmonics behave as
\be
P_{\spin \ell m}(x;s) \sim (1+x)^{\delta_1/2} (1-x)^{\delta_2/2}.
\ee
Since Eqs.~\eqref{eq:func_sep_variables} and \eqref{eq:reg_axis_ansatz} imply
\bea
\bar \psi_{\spin  m }(\sigma, x; s) &=& \sum_{\ell=\max(|\spin|,|m|)}^\infty \bar \Phi_{\spin  \ell m }(\sigma; s) P_{\spin  \ell m }(x; s) \nn
&& \qquad \times (1+x)^{\delta_1/2}(1-x)^{\delta_2/2},
\eea
it follows $\bar \psi_{\spin  m }(\sigma, x; s) \sim (1+x)^{\delta_1} (1-x)^{\delta_2}$. In this way, the Ansatz \eqref{eq:reg_axis_ansatz} ensures $\bar \psi_{\spin  m }(\sigma, \pm 1; s) = 0$ whenever $|m|\neq |\spin|$. For $|m| = |\spin|$, $\bar \psi_{\spin  m }(\sigma, x; s)$ vanishes either at $x=1$ or $x=-1$ (or neither if $\spin = 0$).

\paragraph*{Pole conditions.}
The vanishing of $\bar\psi_{\spin m}$ at the poles is \emph{not} enforced by the field equation alone. Near $x=-1$, with $u=1+x$, the leading part of Eq.~\eqref{eq:aux_field} is $2u\,\partial^2_u\bar\psi + 2(1-\delta_1)\partial_u\bar\psi = 0$, with Frobenius exponents $0$ and $\delta_1$; at $x=+1$ the exponents are $0$ and $\delta_2$. For integer $\delta_{1,2}>0$ both branches are smooth, so a spectral discretisation that imposes the field equation at the axis admits the branch $\bar\psi\sim(1\pm x)^0$, which corresponds to the irregular solution $\bar\Phi\sim(1\pm x)^{-\delta_{1,2}/2}$ of the original problem. Whenever $\delta_1>0$ ($\delta_2>0$) we therefore impose the Dirichlet condition $\bar\psi_{\spin m}=0$ at $x=-1$ ($x=+1$), see Sec.~\ref{sec:BCs}. Without it the discrete operator has an (exact or near) null space and the numerical solution does not converge with resolution. For $\delta_1=0$ ($\delta_2=0$), instead, the two exponents coincide and the second branch is logarithmic, $\bar\psi\sim\log(1\pm x)$. It is excluded by the spectral representation, and the field equation alone selects the regular solution, which does not vanish at the pole.

The different behaviour at $x=1$ and $x=-1$ for $|m| = |\spin|$ points to a break of symmetry across the equator under $x\to -x$. The next section discusses this property in more detail.

\subsubsection{Symmetry across the equator \texorpdfstring{$x=0$}{x=0}}

When solving Eq.~\eqref{eq:aux_field} for the scalar field ($\spin=0$) on Schwarzschild ($\kappa=0$), Ref.~\cite{PanossoMacedo:2024pox} exploited the equatorial symmetry to reduce the domain of dependence for the angular variable to $x\in[0,1]$. Indeed, for $(\spin, \kappa)=(0,0)$ and any $m$ value, the coefficients \eqref{eq:ang_gxx}-\eqref{eq:ang_g0} satisfy ${\frak g}_{xx}(-x)={\frak g}_{xx}(x)$, ${\frak g}_{0}(-x)={\frak g}_{0}(x)$. Besides, ${\frak g}_{x}(-x)=-{\frak g}_{x}(x)$, which ensures the equatorial symmetry together with $\partial_x \to -\partial_x$ under $x\to -x$. The same properties hold for the configuration $(m, \kappa)=(0,0)$ and any $\spin$ value. Hence, equatorial symmetry exists only in specific configurations restricted to the Schwarzschild background.

Since we want to exploit the multi-domain decomposition introduced in Ref.~\cite{PanossoMacedo:2024pox} -- see Sec.~\ref{sec:mdsm} -- we now introduce auxiliary fields with a well-defined symmetry across the equatorial plane.

Recall that any function $f(x)$ for $x\in[-1,1]$ can be split into an even and odd piece $f(x) = f^{\rm E}(x) + f^{\rm O}(x)$, with $f^{\rm E}(-x)=f^{\rm E}(x)$ and $f^{\rm O}(-x)=-f^{\rm O}(x)$. This property is easily seen by considering the Taylor expansion of $f(x)$ around $x=0$, and gathering the even powers of $x$ into $f^{\rm E}(x)$ and the odd into $f^{\rm O}(x)$. Moreover, the odd piece can be further manipulated into $f^{\rm O}(x) = x \check f^{\rm O}(x)$, with $\check f^{\rm O}(-x)=\check f^{\rm O}(x)$.

Considering the above arguments, we split the hyperboloidal master function and source term into
\begin{align}
\label{eq:master_Func_even_odd}
\bar \psi_{\spin  m }(\sigma, x; s) ={}& \bar \psi_{\spin  m }^{\rm E}(\sigma, x; s) + x\, \bar \psi_{\spin  m }^{\rm O}(\sigma, x; s) \\
\label{eq:Source_even_odd}
\bar{\frak S}_{\spin  m }(\sigma, x; s) ={}& \bar{\frak S}_{\spin  m }^{\rm E}(\sigma, x; s) + x\, \bar{\frak S}_{\spin  m }^{\rm O}(\sigma, x; s),
\end{align}
The individual equations for $\bar \psi_{\spin  m }^{\rm E}(\sigma, x; s)$ and $\bar \psi_{\spin  m }^{\rm O}(\sigma, x; s)$ follow from Eq.~\eqref{eq:aux_field} by combining contributions from the sphere's north $x>0$
\be
\left(\bar {\frak D}\bar \psi\right)_{\spin m s}^{\rm N} := \bar {\frak D}_{\spin m s}\,\big[ \bar \psi_{\spin  m}(\sigma, x;s)\big],
\ee
and south hemispheres $x<0$
\be
\left(\bar {\frak D}\bar \psi\right)_{\spin m s}^{\rm S}:=\left.\left(\bar {\frak D}\bar \psi\right)_{\spin m s}^{\rm N}\right|_{x\to -x}
\ee
into the equations of motion
\begin{align}
    \dfrac{\left(\bar {\frak D}\bar \psi\right)_{\spin m s}^{\rm N} + \left(\bar {\frak D}\bar \psi\right)_{\spin m s}^{\rm S}}{2} ={}& \bar {\frak S}^{\rm E}_{\spin m}(\sigma, x;s),\label{eq:even_eq_calc}\\
    \dfrac{\left(\bar {\frak D}\bar \psi\right)_{\spin m s}^{\rm N} - \left(\bar {\frak D}\bar \psi\right)_{\spin m s}^{\rm S}}{2 x} ={}& \bar {\frak S}^{\rm O}_{\spin m}(\sigma, x;s). \label{eq:odd_eq_calc}
\end{align}

Recall that the extra factor of $x$ in the {\em odd} part of Eqs. \eqref{eq:master_Func_even_odd} and \eqref{eq:Source_even_odd} enforces $\bar \psi_{\spin  m }^{\rm O}(\sigma, x; s)$ and $\bar{\frak S}_{\spin  m }^{\rm O}(\sigma, x; s)$ to be even functions, which allows us to restrict the angular domain to $x\in[0,1]$. Keeping this property in mind, we will abuse the language and still refer to $\bar \psi_{\spin  m }^{\rm O}(\sigma, x; s)$ -- also $\bar{\frak S}_{\spin  m }^{\rm O}(\sigma, x; s)$ -- as the ``odd'' field. The values on the southern hemisphere are recovered from $\bar\psi_{\spin m}(\sigma,-x) = \bar\psi^{\rm E}_{\spin m} - x\,\bar\psi^{\rm O}_{\spin m}$; in particular the north and south poles correspond to $\bar\psi^{\rm E}\pm\bar\psi^{\rm O}$ at $y=1$.

Since the Taylor expansion of ``even'' and ``odd'' fields contains only powers of $x^2$, we introduce a final transformation
\be
\label{eq:x_to_y}
y = x^2, \quad y \in [0,1],
\ee
to properly account for the regularity condition at the equatorial plane $y=0$~\cite{PanossoMacedo:2024pox}.

By explicitly evaluating the left-hand side of Eqs.~ \eqref{eq:even_eq_calc}--\eqref{eq:odd_eq_calc}, and incorporating the coordinate change \eqref{eq:x_to_y}, we obtain the equations coupling $\bar \psi_{\spin  m }^{\rm E}(\sigma, y; s)$ and $\bar \psi_{\spin  m }^{\rm O}(\sigma, y; s)$
\begin{align}
    \hat {\bar {\frak D}}^{\rm E}_{\spin m s} \big[\vec {\bar \psi}_{\spin  m}(\sigma, y;s)  \big] ={}& \bar {\frak S}^{\rm E}_{\spin m}(\sigma, y;s), \\
    \hat {\bar {\frak D}}^{\rm O}_{\spin m s} \big[\vec {\bar \psi}_{\spin  m}(\sigma, y;s)  \big] ={}& \bar {\frak S}^{\rm O}_{\spin m}(\sigma, y;s)
\end{align}
with the state vector
\be
\vec {\bar \psi}_{\spin  m}(\sigma, y;s) = \left(
\begin{array}{c}
\bar \psi_{\spin  m }^{\rm E}(\sigma, y; s) \\
\bar \psi_{\spin  m }^{\rm O}(\sigma, y; s)
\end{array}
\right)
\ee
and the even/odd operators (acting as row vectors on $\vec{\bar\psi}$)
\bea
&& \hat {\bar {\frak D}}^{\rm E}_{\spin m s} = \left(
\bar{\cal R}_{\spin m s} + \bar{\frak X}^{\rm E}_{\spin m s}\,, \quad \bar{\frak C}^{\rm EO}_{\spin m s}
\right), \\
&& \hat {\bar {\frak D}}^{\rm O}_{\spin m s} = \left(
 \bar{\frak C}^{\rm OE}_{\spin m s}\,, \quad \bar{\cal R}_{\spin m s} + \bar{\frak X}^{\rm O}_{\spin m s}
\right).
\eea
While the operator acting on the radial direction remains the same along all the previous manipulations,  cf.~Eqs.~\eqref{eq:Radial_Operator} and \eqref{eq:radial_op_functions}, the angular operators take the final form
\begin{align}
    \bar{\frak X}^{\rm E}_{\spin m s} ={}& 4 y (1-y)\partial^2_{yy} + 2\big[1 - y(3 - \delta_1 - \delta_2) \big]\partial_{y} \nn
        & - \dfrac{m^2 - \delta_1 - \delta_2 + \delta_1 \delta_2 - \spin(2+\spin) }{2} \nn
        & + \dfrac{\rh^2 s^2\kappa^2}{\lambda^2}(1-y) , \\
    \bar{\frak X}^{\rm O}_{\spin m s} ={}& 4 y (1-y)\partial^2_{yy} + 2\big[3 - y( 5 - \delta_1 - \delta_2) \big]\partial_{y} \nn
        & - \dfrac{4 + m^2 - 3(\delta_1+ \delta_2) + \delta_1 \delta_2 - \spin(2+\spin) }{2} \nn
        & + \dfrac{\rh^2 s^2\kappa^2}{\lambda^2}(1-y),
\end{align}
and the coupling operators
\begin{align}
    \bar{\frak C}^{\rm EO}_{\spin m s}  ={}& {-}2 y (\delta_1 -\delta_2)\partial_y - \delta_1 +\delta_2 - \dfrac{2 i \rh s \kappa \spin}{\lambda} y, \\
    \bar{\frak C}^{\rm OE}_{\spin m s}  ={}& {-}2 (\delta_1 -\delta_2)\partial_y - \dfrac{2 i \rh s \kappa \spin}{\lambda}.
\end{align}
The even and odd fields decouple when $\delta_1=\delta_2$ and $\kappa\,\spin=0$, which recovers the equatorial symmetry discussed above.

\paragraph*{Regularity at the equator.} The point $y=0$ is a regular singular point of $\bar{\frak X}^{\rm E}$ and $\bar{\frak X}^{\rm O}$, with Frobenius exponents $\{0,\tfrac12\}$ and $\{0,-\tfrac12\}$, respectively. The non-zero exponents correspond to non-smooth functions of $x$ ($|x|$ and $1/|x|$), and are excluded by the spectral representation in $y$. Hence, as in Ref.~\cite{PanossoMacedo:2024pox}, the field equations are imposed directly at $y=0$, without additional conditions.

\subsection{Regularity along the radial direction}\label{sec:radial_reg}

The radial operator~\eqref{eq:Radial_Operator} is degenerate at both ends of $\sigma\in[0,1]$, since $p(\sigma)=\sigma^2(1-\sigma)(1-\sigma/\sigma_{\rm c})$ vanishes there. The physical (retarded) solution is selected by demanding smoothness, and no boundary data are imposed. For radial fixing and $\lambda=\rh$:

\subsubsection{Future null infinity \texorpdfstring{$\sigma=0$}{σ=0}}
Here ${\rm R}_\sigma(0) = 2s$, so near $\sigma=0$ the radial equation reduces to $\sigma^2\partial_\sigma^2\bar\psi + 2s\,\partial_\sigma\bar\psi \simeq 0$. Besides a regular solution, it admits a branch with $\partial_\sigma\bar\psi \propto e^{2s/\sigma}$, i.e.\ an essential singularity which, for $s=-i\lambda\omega$, oscillates infinitely fast as $\sigma\to0$ and corresponds to radiation incoming from past null infinity~\cite{Minucci:2024qrn}. It cannot be represented by a polynomial, and imposing the field equation at $\sigma=0$ selects the outgoing solution.

For the static mode, $s=0$, the essential singularity is absent. Near $\sigma=0$ one then has ${\rm R}_{\sigma\sigma}\simeq\sigma^2$, ${\rm R}_\sigma\simeq2(1+\spin)\sigma$ and ${\rm R}_0={\cal O}(\sigma)$, while the angular operator $\bar\Xi_{\spin m s}$ reduces to that of the spin-weighted spherical harmonics, with eigenvalues $-A_{\spin\ell}=-(\ell-\spin)(\ell+\spin+1)$. For each $\ell$-mode, $\sigma=0$ is therefore a regular singular point, with indicial equation $k^2+(1+2\spin)k-A_{\spin\ell}=0$ and Frobenius exponents $k=\ell-\spin$ and $k=-(\ell+\spin+1)$. Since $\ell\ge|\spin|$, the second exponent is negative and the corresponding branch diverges at $\scri^+$, so imposing the field equation at $\sigma=0$ again selects the regular solution, which for $\spin=-2$ vanishes as $\sigma^{\ell+2}$.

\subsubsection{Black hole horizon \texorpdfstring{$\sigma=1$}{σ=1}}
Near $\sigma=1$, with $u=1-\sigma$, one has $p\simeq(1-\kappa^2)\,u$ and the radial equation becomes $(1-\kappa^2)u\,\partial_u^2\bar\psi - {\rm R}_\sigma(1)\,\partial_u\bar\psi\simeq0$, with Frobenius exponents $0$ and
\be
\nu = 1+\dfrac{{\rm R}_\sigma(1)}{1-\kappa^2} = \spin + \dfrac{i\,(\omega - m\Omega_{\rm h})}{\varkappa_{\rm h}}.
\ee
The branch $(1-\sigma)^\nu$ is the wave outgoing from the horizon~\cite{Minucci:2024qrn}. For $\spin=-2$ it is singular (or, for $\omega=m\Omega_{\rm h}$, a negative integer power), so imposing the field equation at $\sigma=1$ again selects the physical, ingoing solution. This holds also for the static mode $m=0$, for which $\nu=\spin$.

\section{Secondary black hole}\label{sec:particle}

\subsection{Particle's trajectory}
Owing to the preservation of time-translation symmetry around a Kerr black hole, the geodesic motion of a test mass (in this case, the secondary black hole) is characterised by the two conventionally conserved orbital quantities, the total energy per unit mass $E$ and the projection of the orbital angular momentum per unit mass along the symmetry axis, $L_z$, as well as the Carter constant, $Q$. The first two quantities arise from the constants of motion associated with the Killing vectors ${\xi_{(t)}^a = \partial_t}$ and $\xi_{(\phi)}^a = \partial_\phi$, with
\begin{align}
	E = -\xi_{(t)}^a u_a \, , \quad L_z = \xi_{(\phi)}^a u_a
\end{align}
where $u^a$ is the particle's four-velocity.

We consider a particle on a circular, equatorial orbit $x_p(\uptau) = (t(\uptau), r_0, \pi/2, \phi_p(\uptau))$, with $\uptau$ the particle's proper time. We allow the black-hole spin parameter to take both signs, $a/M\in[-1,1]$, and orient the azimuthal coordinate such that the particle always moves towards increasing $\phi$. The particle's four-velocity then reads
\bea
\label{eq:partc_velocity}
&& u^\mu = u^t \left( \delta^\mu_t + \Omega_\phi \, \delta^\mu_\phi\right), \quad \Omega_\phi = \dfrac{1}{\sqrt{r_0^3/M} + a}, \\
&& u^t = \dfrac{\sqrt{r_0^3/M} + a}{\sqrt{r_0^3/M - 3 r_0^2 + 2a \sqrt{r_0^3/M}}},
\eea
and the azimuthal angle is parametrised by $\phi_p(t) = \Omega_\phi t$, with $\Omega_\phi>0$.

In this convention, the sign of $a$ distinguishes the two orbital orientations. For $a>0$ the black hole rotates in the same sense as the particle (prograde orbits), whereas for $a<0$ it rotates in the opposite sense (retrograde orbits). All spin-dependent quantities, such as $\kappa=a/\rh$, $\Omega_{\rm h}$ and $\varphi_*(\sigma)$, carry the sign of $a$. In particular, the superradiance condition $\omega_m<m\,\Omega_{\rm h}$ can only be met by prograde orbits.

The more common convention restricts $a\geq0$ and describes retrograde orbits by $\Omega_\phi<0$. It is related to ours by the reflection $\phi\to-\phi$, which maps $a\to-a$ and $\Omega_\phi\to-\Omega_\phi$. Since this reflection also maps $m\to-m$, the mode $m$ of a retrograde orbit with $a<0$ corresponds to the mode $-m$ in the $a\geq0$ convention, whereas quantities summed over $\pm m$, such as the total fluxes, are identical in both conventions. 

\subsection{Comoving frame}

Following Ref.~\cite{lmtompaper}, where full details and the derivation can be found, we introduce a locally Cartesian coordinate system $X^\mu = \{T,X,Y,Z\}$ around the secondary particle's worldline $x^\mu(\uptau)=x^\mu_p$ that is comoving with the particle.
With respect to Boyer--Lindquist coordinates, the comoving coordinates are
\begin{align}
\label{eq:comoving_coords}
    T ={}& E t - L_z \varphi, \\
	X ={}& \dfrac{r-r_0}{\mathscr B}, \\
    Y ={}& {-} r_0 \cos \theta, \\
    Z ={}& z_c \sin\biggl(\dfrac{\hat{\varphi}}{2}\biggr)
\end{align}
where $z_c \coloneqq 2u^{t}r_0\,\mathscr B$, $\hat{\varphi} \coloneqq \varphi - \varphi_p$, and
\bea
\label{eq:B_def}
\mathscr B \coloneqq \sqrt{\bar\Delta(\sigma_0)} &=& \sqrt{(1-\sigma_0)(1-\kappa^2\sigma_0)} \\
&=& \sqrt{1-\dfrac{2M}{r_0} + \dfrac{a^2}{r_0^2}} . \notag
\eea
Here, $\bar\Delta=\Omega^2\Delta$ is the conformally rescaled function of Eq.~\eqref{eq:conf_KerrFunc}, evaluated at the particle's position $\sigma_0=\rh/r_0$.\footnote{Eq.~\eqref{eq:B_def} assumes radial fixing, $\rho_o=0$, with $\lambda=\rh$, for which $\Omega(\sigma_0)=1/r_0$ and hence $\mathscr B=\sqrt{\Delta(r_0)}/r_0$. For other choices, $\mathscr B$ is defined directly by the latter expression.}
These coordinates are used both in the construction of the puncture and effective source (based on that of Ref.~\cite{lmtompaper}) and they also define the shape of the particle domain, see Sec.~\ref{sec:mdsm}.

\subsection{Puncture field and effective source}\label{sec:puncture}

In the puncture scheme, we begin by splitting the metric perturbation into two pieces:
\begin{equation}
    h^{(1)}_{ab} = h^{\calP1}_{ab} + h^{\calR1}_{ab}. \label{eq:h1PR}
\end{equation}
The puncture field, $h^{\calP1}_{ab}$, is an analytically known expression that captures the singularity structure of the small object and is a local expansion near the worldline of the object's Detweiler--Whiting singular field~\cite{Detweiler:2002mi}.
The residual field, $h^{\calR1}_{ab}$, then determines the object's motion and whose smoothness is dictated by the expansion order of the puncture.
In this work, we use the punctures derived in Ref.~\cite{lmtompaper} which means that $h^{\calR1}_{ab}$ is $C^2$ on the worldline; see Ref.~\cite{lmtompaper} for further details.

The puncture field is then written in the comoving coordinates $(T,X,Y,Z)$ of the previous subsection with the form of a polynomial in $X$, $Y$ and $Z$ multiplied by an odd power of $R$ with coefficients depending on $M$, $r_0$ and $a$, and where the sum of all powers is between $-1$ and $2$.
Its $m$-mode decomposition is given as an integral over $\hat{\varphi}$ at fixed $(X,Y)$ which is carried out analytically.
The resulting $m$-mode puncture is a closed-form function of $\rho=\sqrt{X^2+Y^2}$ and of the variable $z=(\rho^2+z_c^2/2)/(\rho\sqrt{\rho^2+z_c^2})$, through associated Legendre functions $P^{i}_{j/2}(z)$, where the values of $i$ and $j$ are dictated by the powers of $Z$ and $R$.

To construct the Weyl scalar puncture, we follow Eq.~\eqref{eq:Th1} so that
\begin{equation}
    \psi^{\rm P}_m = \That_{4}[h^{\calP1}_m].
\end{equation}
From this, the effective source is obtained by applying the Teukolsky operator~\eqref{eq:Teuk_Op} analytically to the puncture.
Both are evaluated in Boyer--Lindquist coordinates and then transformed to the hyperboloidal variables discussed above before finally being split into even/odd components, as in Sec.~\ref{sec:Hyp_Teuk}.

Within the particle domain $\mathcal D_2$ we solve for the residual field $\bar\psi^{\rm R}_{\spin m} = \bar\psi_{\spin m}-\bar\psi^{\rm P}_{\spin m}$, which satisfies
\be
\bar {\frak D}_{\spin ms}\,\bar\psi^{\rm R}_{\spin m} = \bar{\frak S}^{\rm eff}_{\spin m} := -\,\bar {\frak D}_{\spin ms}\,\bar\psi^{\rm P}_{\spin m},
\ee
where $\bar{\frak S}^{\rm eff}_{\spin m}$ is the effective source and the form of the operator $\bar{\frak D}_{\spin ms}$ is given in Sec.~\ref{sec:reg_ang}. With the order of puncture being used, the effective source is logarithmically divergent at the particle meaning that $\bar\psi^{\rm R}_{\spin m}$ is $C^1$ there (Sec.~\ref{sec:results}).
From Sec.~\ref{sec:mdsm} onwards, we omit the indices $\spin m$ on the axis-regularised field $\bar\psi_{\spin m}$, on its residual and puncture parts, $\bar\psi^{\rm R}$ and $\bar\psi^{\rm P}$, and on the effective source $\bar{\frak S}^{\rm eff}$.

\section{Multi-domain square mapping}\label{sec:mdsm}

\subsection{Even/Odd Decomposition}
As described in Sec.~\ref{sec:axis}, the field is decomposed into an even field $\bar\psi^{\rm E}(\sigma, y)$ and an ``odd'' field $\bar\psi^{\rm O}(\sigma, y)$, both functions of $y=x^2$, such that
\begin{align}
	\bar\psi(\sigma,x) = \bar\psi^{\rm E}(\sigma, y) + x \,\bar\psi^{\rm O}(\sigma, y) \, .
\end{align}
Even though this doubles the number of fields to solve for, it allows us to restrict the domain of $\theta$ to the northern hemisphere $\theta\in[0, \pi/2]$, while still describing solutions without equatorial symmetry.

\subsection{Domains}
\begin{figure}
    \centering
    \includegraphics[width=\linewidth]{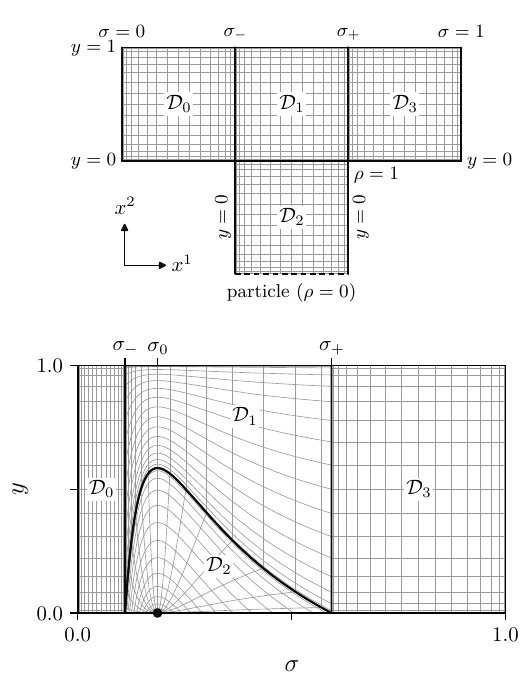}
    \caption{Schematic of the mutlidomain scheme. \emph{Above:} the domain decomposition in terms of the spectral coordinates $(x^1,x^2)$ which run from $[-1,1]$ in each $\mathcal D_i$. Note that the $y=0$ surface continues along the outer edges of $\mathcal D_2$. The particle position $\rho=(1+x^2)/2 = 0$ is excluded in the Radau grid and so is denoted by a dashed line. \emph{Below:} the domain mapping in terms of $(\sigma,y)$ for $a=0.5$, $r_0=10M$. Grid lines denote constant $x^1$ or $x^2$ in each domain.}
    \label{fig:domain_schematic}
\end{figure}

Following the multidomain scheme outlined in~\cite{PanossoMacedo:2024pox}, we divide the region $(\sigma,y)\in[0,1]^2$ into the four domains $\mathcal{D}_i$ shown in Fig.~\ref{fig:domain_schematic} 
is for fixed worldtube size and only demonstrates the effects of $|a|$ and $r_0$ on the geometry.
See figure~\ref{fig:domain_schematic} for generic mapping.
Within each domain, we choose mappings that take the hyperboloidal coordinates ${(\sigma, y)}$ to a square region ${(x^1, x^2) \in [-1,1]^2}$.
With $\sigma_0 = \rh/r_0$ the particle's position, we define
\be
\label{eq:world_tube_bounds}
\sigma_+ = \dfrac{1+\sigma_0}{2}, \quad \sigma_- = \sigma_0\,\sigma_+.
\ee
This definition follows the strategy of Ref.~\cite{PanossoMacedo:2024pox}, setting the inner boundary of the worldtube, $\sigma_+$, halfway (in $\sigma$) between the particle and the horizon. The outer boundary $\sigma_-$ then follows from requiring the worldtube to be a disc in the comoving coordinates. Appendix~\ref{app:worldtube} discusses the alternative where the world-tube is re-dimensioned.

%

In more detail:\\

\textbf{Domain 0: Outer source-free region,} $(r,\theta) \in [r_+, \infty) \times [0,\pi/2]$, where $r_+=\rh/\sigma_->r_0$. The map reads
\begin{align}
	\sigma &= \sigma_- (1 + x^1)/2, \quad
    y = (1 + x^2)/2 .
\end{align}

\textbf{Domain 3: Inner source-free region,} $(r,\theta) \in [r_h, r_-] \times [0,\pi/2]$, where $r_-=\rh/\sigma_+<r_0$. The map reads
\begin{align}
	\sigma &= \sigma_+ \dfrac{1 - x^1}{2} + \dfrac{1+x^1}{2}, \quad
    y = (1 + x^2)/2 .
\end{align}

\textbf{Domain 2: Effective source region.} This region describes a disc of radius $R_0 = \eta\,\rh$ centred at the particle in the comoving coordinates $(X,Y)$, which touches domains $\mathcal{D}_0$ and $\mathcal{D}_3$ at the equator. We introduce polar coordinates $\{\rho, \bar\vartheta\}$ centred around the particle, $X = -R_0\,\rho\cos\bar\vartheta$, $|Y| = R_0\,\rho\sin\bar\vartheta$, such that this domain corresponds to $(\rho,\bar\vartheta) \in [0, 1] \times [0,\pi]$. They map to the square via
\begin{equation}
    x^1=\cos\bar\vartheta\,,\quad x^2=2\rho-1,
\end{equation}
and, in terms of the hyperboloidal coordinates,
\begin{align}
\label{eq:comovingDisk_hypMap}
\sigma &= \dfrac{\sigma_0}{1 - \eta\,\mathscr{B}\,\sigma_0\,\rho\, x^1}, \quad
y = \eta^2\sigma_0^2\,\rho^2\left[1-(x^1)^2\right] .
\end{align}
The disc touches $\mathcal{D}_0$ and $\mathcal{D}_3$ at $\rho=1$, $x^1=\mp1$, where Eq.~\eqref{eq:comovingDisk_hypMap} gives
\begin{equation}
\sigma_\pm=\dfrac{\sigma_0}{1\mp \eta\,\mathscr B\,\sigma_0} \Longleftrightarrow r_\mp=r_0\mp \eta\,\mathscr B\,\rh .
\end{equation}
Identifying $\sigma_+$ with its value in Eq.~\eqref{eq:world_tube_bounds} fixes the disc radius
\begin{equation}
\eta = \dfrac{1-\sigma_0}{\sigma_0\,\mathscr B\,(1+\sigma_0)},
\end{equation}
and consistently reproduces $\sigma_-=\sigma_0\,\sigma_+$.

\textbf{Domain 1: Intermediate source-free region.} This domain covers the part of the shell $\sigma \in [\sigma_-, \sigma_+]$ not covered by domain $\mathcal{D}_2$. Mapping this to a square domain $\{x^1,x^2\} \in [-1,1]^2$ gives
\begin{align}
\sigma&=\sigma_0(x^1) \coloneqq \dfrac{\sigma_0}{1 - \eta\,\mathscr{B}\,\sigma_0\, x^1},\nn
y&=y_0(x^1)\dfrac{1-x^2}{2}+\dfrac{1+x^2}{2}\,,
\end{align}
where $y_0(x^1)=\eta^2\sigma_0^2[1-(x^1)^2]$. The curve $(\sigma_0(x^1),y_0(x^1))$ is the outer boundary of domain $\mathcal{D}_2$ ($\rho=1$); $\mathcal D_1$ and $\mathcal D_2$ therefore share the same coordinate $x^1$ along their interface.

\subsection{Boundary conditions}\label{sec:BCs}

The conditions imposed on the edges of each domain are of three kinds: (i) transition conditions at the interfaces between domains; (ii) the field equations themselves, at boundaries of $[0,1]^2$ where the principal part of the operator degenerates (regularity conditions); and (iii) Dirichlet conditions at the poles. They are summarised in Table~\ref{tab:BCs}.

\paragraph*{Interfaces.} In the source-free domains $\mathcal{D}_0, \mathcal{D}_1, \mathcal{D}_3$ we solve directly for the retarded field. Across the interfaces $\mathcal{D}_0-\mathcal{D}_1$ and $\mathcal{D}_1-\mathcal{D}_3$ we impose continuity of the solution and of its first $\sigma$-derivative. In the effective source region, we instead solve for the residual field. Continuity of the retarded field and of its derivative normal to the interface $\mathcal{D}_1-\mathcal{D}_2$ then imposes jump conditions on the solution, given by the puncture field and its normal derivative:
 \begin{align}
     &\mathcal{D}_0-\mathcal{D}_1\!\!:&&\hspace{-8pt}\bar\psi^{A,1}\big\rvert_{x^1=-1}-\bar\psi^{A,0}\big\rvert_{x^1=1}=0,\nn
   &&&\hspace{-8pt} \partial_\sigma\bar\psi^{A,1}\big\rvert_{x^1=-1}-\partial_\sigma\bar\psi^{A,0}\big\rvert_{x^1=1}=0,\nn
    &\mathcal{D}_1-\mathcal{D}_3\!\!:&&\hspace{-8pt}\bar\psi^{A,3}\big\rvert_{x^1=-1}-\bar\psi^{A,1}\big\rvert_{x^1=1}=0,\nn
    &&&\hspace{-8pt} \partial_\sigma\bar\psi^{A,3}\big\rvert_{x^1=-1}-\partial_\sigma\bar\psi^{A,1}\big\rvert_{x^1=1}=0,\nn
    &\mathcal{D}_1-\mathcal{D}_2\!\!:&&\hspace{-8pt} \bar\psi^{A,2}\big\rvert_{x^2=1}-\bar\psi^{A,1}\big\rvert_{x^2=-1}=-\bar\psi^{A,{\rm P}}\big\rvert_{x^2=1},\nn
    &&&\hspace{-8pt} \partial_{n_2}\bar\psi^{A,2}\big\rvert_{x^2=1}-\partial_{n_2}\bar\psi^{A,1}\big\rvert_{x^2=-1}=-\partial_{n_2}\bar\psi^{A,{\rm P}}\big\rvert_{x^2=1},
 \end{align}
where $A=\rm E, O$ and $\partial_{n_2}$ is the derivative normal to the curve $x^2=$ const, given by
\begin{equation}
    \partial_{n_2}=\dfrac{\sigma_{,x^1} \partial_y-y_{,x^1} \partial_\sigma}{\sqrt{\sigma_{,x^1}^2+y_{,x^1}^2}} .
\end{equation}
Each pair of conditions is distributed over the two sides of an interface: the continuity of the field is imposed on the collocation points of one domain, and that of the normal derivative on those of the other (Table~\ref{tab:BCs}). Note that the derivatives of the puncture are computed analytically from the puncture expression, and then evaluated numerically on the grid.

\paragraph*{Regularity.} At future null infinity ($\sigma=0$), the horizon ($\sigma=1$) and the equator ($y=0$), the operator degenerates and the field equations themselves select the regular solution (Secs.~\ref{sec:axis} and \ref{sec:radial_reg}). At these boundaries we therefore impose the field equations, external without boundary data~\cite{PanossoMacedo:2024pox}. 

\paragraph*{Pole conditions.} As shown in Sec.~\ref{sec:axis}, the field equation at the axis does not exclude the irregular branch $\bar\psi\sim(1\pm x)^0$. In terms of the even/odd fields, the north ($x=1$) and south ($x=-1$) poles both lie at $y=1$, where $\bar\psi = \bar\psi^{\rm E}\pm\bar\psi^{\rm O}$. On the axis points of $\mathcal D_0$, $\mathcal D_1$ and $\mathcal D_3$ we replace
\begin{itemize}
\item the odd field equation by $\bar\psi^{\rm E}-\bar\psi^{\rm O}=0$ (south pole) if $\delta_1>0$, and keep the even field equation;
\item the even field equation by $\bar\psi^{\rm E}+\bar\psi^{\rm O}=0$ (north pole) if $\delta_2>0$, and keep the odd field equation.
\end{itemize}
Since the original even/odd field equations also degenerate at the axis, they act as the complementary regularity condition and, together with the substitutions, the system is closed. When $\delta_1=0$ ($\delta_2=0$), no south-pole (north-pole) condition is needed (Sec.~\ref{sec:axis}), and the odd (even) field equation is imposed without substitution.
For $\spin=-2$ and $m=2$, for example, only the south-pole condition is imposed ($\delta_1=4$, $\delta_2=0$), and the solution at the north pole is non-zero (Fig.~\ref{fig:a0_r10_m2_N32}). 

\paragraph*{Particle domain.} In $\mathcal D_2$, the edges $x^1=\pm1$ lie on the equator, $y=0$, where, as in the other domains, the field equations are imposed as regularity conditions. The radial coordinate $\rho$ extends down to the particle, $\rho=0$, where regularity conditions on the residual field would in principle also be required. Their explicit derivation is cumbersome and, as in Ref.~\cite{PanossoMacedo:2024pox}, can be avoided by a suitable choice of collocation points (Sec.~\ref{sec:grids}).

\begin{table}[t]
\centering
\setlength{\tabcolsep}{4pt}
\begin{tabular}{lll}
\hline\hline
Domain & Edge & Condition imposed \\
\hline
$\mathcal D_0$ & $x^1=-1$ ($\sigma=0$) & field equations \\
 & $x^1=1$ ($\sigma=\sigma_-$) & $\partial_\sigma$-continuity with $\mathcal D_1$ \\
 & $x^2=-1$ ($y=0$) & field equations \\
 & $x^2=1$ ($y=1$) & field eqs.\ / pole conditions \\
$\mathcal D_1$ & $x^1=-1$ ($\sigma=\sigma_-$) & field continuity with $\mathcal D_0$ \\
 & $x^1=1$ ($\sigma=\sigma_+$) & $\partial_\sigma$-continuity with $\mathcal D_3$ \\
 & $x^2=-1$ ($\rho=1$) & $\partial_{n_2}$-jump with $\mathcal D_2$ \\
 & $x^2=1$ ($y=1$) & field eqs.\ / pole conditions \\
$\mathcal D_2$ & $x^1=\pm1$ ($y=0$) & field equations \\
 & $x^2=-1$ ($\rho\gtrsim0$) & field equations \\
 & $x^2=1$ ($\rho=1$) & field jump with $\mathcal D_1$ \\
$\mathcal D_3$ & $x^1=-1$ ($\sigma=\sigma_+$) & field continuity with $\mathcal D_1$ \\
 & $x^1=1$ ($\sigma=1$) & field equations \\
 & $x^2=-1$ ($y=0$) & field equations \\
 & $x^2=1$ ($y=1$) & field eqs.\ / pole conditions \\
\hline\hline
\end{tabular}
\caption{Conditions imposed on the collocation points of each domain edge, for each of the four real fields. Interior points carry the field equations. ``Pole conditions'' replace the even (odd) field equation by $\bar\psi^{\rm E}+\bar\psi^{\rm O}=0$ ($\bar\psi^{\rm E}-\bar\psi^{\rm O}=0$) when $\delta_2>0$ ($\delta_1>0$). At a corner, interface conditions take precedence over pole conditions, and pole conditions over field equations; at the two corners of $\mathcal{D}_1$ on $x^2 = -1$, the jump conditions with $\mathcal D_2$ take precedence. There, three domains meet and one matching condition \emph{cannot} be imposed: continuity with $\mathcal D_0$ at $\sigma = \sigma_-$ and $\partial_\sigma$--continuity with $\mathcal D_3$ at $\sigma = \sigma_+$. We have verified numerically that both are satisfied to truncation accuracy.}
\label{tab:BCs}
\end{table}


 \section{Spectral Method}\label{sec:spectral}

We solve the resulting four coupled elliptic partial differential equations with a multi-domain, spectral code, built as an extension of~\cite{PanossoMacedo:2024pox}.  This approach combines the exponential convergence properties of spectral expansions with the flexibility of domain decomposition, allowing us to accurately capture asymptotic behavior at large distances.

\subsection{Spectral Solver}
In this paper we consider $n_d=4$ domains. In each domain, one must solve for $n_f = 4$ real fields, representing the real and imaginary parts of the even and odd parts of $\bar\psi_{-2,m}$. The solution for each field in each domain is approximated using a truncated Chebyshev expansion
\begin{equation}
f^{(i_f,i_d)}_{N^{i_d}_1,\,N^{i_d}_2}(x^1, x^2)
=
\sum_{k_1=0}^{N^{i_d}_1}
\sum_{k_2=0}^{N^{i_d}_2}
c^{(i_d,i_f)}_{k_1,k_2}
\, T_{k_1}(x^1)\, T_{k_2}(x^2)\,,
\end{equation}
where $T_k(x) = \cos(k x)$ are the Chebyshev polynomials of first kind and $N^{i_d}_1,N^{i_d}_2$ correspond, respectively, to the truncation order in the directions $x^1,x^2$ in each domain. All the information about the fields is then included in the Chebyshev coefficients $c^{(i_d,i_f)}_{k_1,k_2}$ which are uniquely fixed by demanding that the full solution coincides with the discretised approximation at the collocation points $\{x^1_{i_1}\}$, $\{x^2_{i_2}\}$, namely
\begin{equation}
f^{(i_f,i_d)}_{i_1,\,i_2}=f^{(i_f,i_d)}_{N^{i_d}_1,\,N^{i_d}_2}(x^1_{i_1}, x^2_{i_2})\,.
\end{equation}
In all results below we use the same truncation $N^{i_d}_1=N^{i_d}_2=N$ in every domain.

Spectral derivatives along the direction $x^i$ are approximated by applying the spectral differentiation matrices, $D^{x^i}_{i_1,j_1}$, on the function approximations $f^{(i_f,i_d)}_{i_1,\,i_2}$. For more information see Appendix C of~\cite{PanossoMacedo:2024pox}.

This discretisation approach reduces both the partial differential equations to be solved and the boundary/transition conditions into a set of linear algebraic equations for the unknowns $f^{(i_f,i_d)}_{i_1,\,i_2}$ that takes the form
\begin{equation}
\label{eq:linear_system}
\hat J \cdot \vec{X}-\vec{S}=0\,,
\end{equation}
where $\vec X$ collects all the $f^{(i_f,i_d)}_{i_1,\,i_2}$ and is of dimension $n_{\rm total}=n_f \sum_{i_d=0}^{n_d-1} n_{1,i_d}\, n_{2,i_d}$, with $n_{j,i_d}=N^{i_d}_j+1$ the number of collocation points along $x^j$ in domain $i_d$.

The matrix $\hat J$ contains the discrete representation of the operator $\hat{\bar{\frak D}}$ and the left-hand sides of the transition and pole conditions of Sec.~\ref{sec:BCs}. The non-vanishing components of the source vector $\vec{S}$ collect the right-hand sides, i.e.\ the effective source in $\mathcal D_2$ and the puncture field and its normal derivative at the $\mathcal D_1-\mathcal D_2$ interface. Writing the set of algebraic equations in the form~\eqref{eq:linear_system}, we have implicitly chosen a specific way for ordering the unknowns into a column vector. Even though the specific choice is not important, this needs to be done consistently for all components of Eq.~\eqref{eq:linear_system}. For completeness, we have used the sequential indexing
\begin{equation}
\label{eq:indexing_func}
I(i_d,i_f,i_1,i_2)=i_f+n_f\Big(i_1+i_2 \,n_{1,i_d}+\sum_{j_d=0}^{i_d-1} n_{1,j_d}n_{2,j_d}\Big),
\end{equation}
where $i_d=0,\ldots,n_d-1$ labels the domain $\mathcal D_{i_d}$ and $i_f=0,\ldots,n_f-1$ the real fields.

The matrix $\hat J$ is assembled analytically in compressed sparse column form directly from the spectral differentiation matrices, and the system is solved with a sparse direct $LU$ factorisation (\texttt{Eigen}~\cite{eigenweb}), followed by one or two steps of iterative refinement with the residual evaluated in quadruple precision (Appendix~\ref{app: performance}). 
Timings are given in Sec.~\ref{sec:timing}.

\subsection{Chebyshev grids}\label{sec:grids}

Following Ref.~\cite{PanossoMacedo:2024pox}, we employ the Chebyshev--Lobatto grid
\be
x^j_{i_j} = \cos\left(\dfrac{\pi i_j}{N}\right), \quad i_j=0,\ldots,N,
\ee
along both directions ($j=1,2$) in $\mathcal D_0$, $\mathcal D_1$, $\mathcal D_3$, and along $x^1$ in $\mathcal D_2$. It includes both end points $x^j=\pm1$, which makes it well suited for imposing boundary and transition conditions. Along $x^2$ in $\mathcal D_2$, we employ the Chebyshev--Radau grid
\be
x^2_{i_2} = \cos\left(\dfrac{2\pi i_2}{2N+1}\right), \quad i_2=0,\ldots,N,
\ee
which includes the interface $\rho=1$ but excludes the particle at $\rho=0$. The field equations imposed at the last node then implicitly enforce regularity at the particle~\cite{PanossoMacedo:2024pox}.

The equatorial edges $x^1=\pm1$ of $\mathcal D_2$ require care: there, the odd part of the effective source $\bar{\frak S}^{\rm eff}$, cf.~Eq.~\eqref{eq:Source_even_odd}, evaluates numerically to $0/0$ despite the formal limit being finite. When attempting to evaluate on a Gauss grid in $x^1$, numerical results suffer from catastrophic cancellation near the equator, spoiling potential extrapolation to the Lobatto grid. In lieu of analytic expressions for the hyperboloidal fields at the equator, we instead employ a cubic extrapolation method at constant $x^2$ to obtain the values\footnote{In particular, we take the points $x^1 = \pm(1-k h)$ for all $x^2$ with $k=1,\ldots,4$ and a fixed $h=4\times10^{-5}$. The error from the extrapolation is $\mathcal{O}(h^4) \sim 10^{-13}$. This method is applied to all the fields (even and odd) as well as their derivatives, and the effective source so that there is a fixed error floor. The equatorial values are appended to the field evaluations on the Lobatto grid and the extrapolation data is discarded.}.


\section{Results}\label{sec:results}

We present results for a particle on a circular equatorial orbit at $r_0=10M$. The main parameter set consists of the spins $a/M=0$, $\pm0.5$, $\pm0.9$, where $a<0$ denotes a retrograde orbit (Sec.~\ref{sec:particle}), and the modes $m=1,\ldots,20$, computed with $N=40$ for $m\le10$ and $N=48$ for $m\ge11$. The higher resolution for $m\ge11$ compensates for the slower convergence of the high-$m$ modes, so that all modes are compared at a similar level of accuracy.
Unlike other numerical infrastructures~\cite{Berndtson:2007gsc,Dolan:2012jg,Dolan:2023enf,lmtompaper, Osburn:2022bby,Nishimura2026}, our code can handle straightforwardly the static mode $m=0$, which is discussed separately in Sec.~\ref{sec:m0}. Unless stated otherwise, the worldtube has its default size as defined in Sec.~\ref{sec:mdsm}. Near-extremal spins, $|a|/M$ up to $0.9999$, other orbital radii ($r_0/M=1.5$, $9$, $15$) and the effect of the size of the worldtube are studied in Appendices~\ref{app: performance} and~\ref{app:worldtube}. There we find that, at fixed $N$, the accuracy decreases as $r_0/\rh$ increases, and that this can be compensated by shrinking the worldtube. The modes with $m<0$ are related to those with $m>0$ by symmetry, and we verified numerically that the corresponding fluxes coincide.

The computational cost of a solve, its scaling with $N$ and the conditioning of the linear system are discussed in Appendix~\ref{app: performance}. Our reference values are computed with the Black Hole Perturbation Toolkit (BHPToolkit)~\cite{BHPToolkit}, using either its Mathematica \texttt{Teukolsky} package or its Python package \texttt{pybhpt}, which solve the separated radial Teukolsky equation mode by mode.

\subsection{Global solution}

\begin{figure*}
    \centering
    \includegraphics[width=0.8\linewidth]{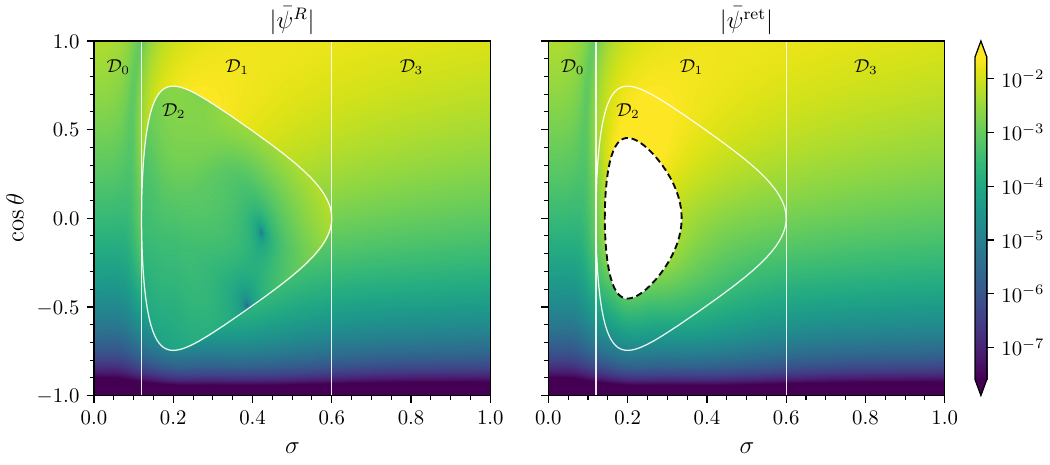}
    \caption{Absolute value of the residual field $\bar\psi^{\rm R}$ and of the retarded field $\bar\psi^{\rm ret}$ for $a/M=0$, $r_0/M=10$, $m=2$, recombined over both hemispheres, $x\in[-1,1]$. The two fields differ only in $\mathcal D_2$, where $\bar\psi^{\rm ret}=\bar\psi^{\rm R}+\bar\psi^{\rm P}$. The residual field jumps across the $\mathcal D_1$--$\mathcal D_2$ interface by the value of the puncture, whereas the retarded field is smooth across all interfaces. This is a non-trivial check of the puncture, the effective source and the jump conditions. For this mode only the south-pole condition is imposed ($\delta_1=4$, $\delta_2=0$), and the field does not vanish at the north pole. Dashed areas mark singular regions near the particle that are not shown, to keep the colour scale readable.}
\label{fig:a0_r10_m2_N32}
\end{figure*}

\begin{figure}
    \centering
    \includegraphics[width=\linewidth]{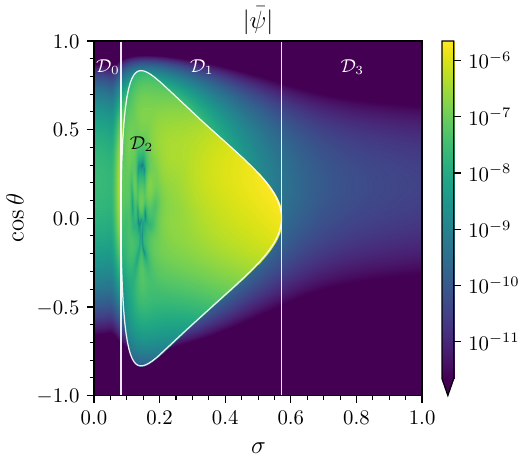}
    \caption{Absolute value of the retarded field $\bar\psi^{\rm ret}$ in $\mathcal{D}_0,~\mathcal{D}_1,~\mathcal{D}_3$ and $\bar\psi^R$ in $\mathcal D_2$ for the retrograde orbit, $a/M=-0.9$, $r_0/M=10$, and $m=10$, over both hemispheres. Both pole conditions are imposed ($\delta_1,\delta_2>0$), and the field vanishes at both poles.}
    \label{fig:other_global_sln}
\end{figure}

Figures~\ref{fig:a0_r10_m2_N32} and \ref{fig:other_global_sln} show the modulous  of regularised field for two representative cases at $r_0=10M$: the dominant mode $m=2$ for a Schwarzschild black hole and the higher mode $m=10$ for a retrograde orbit around a rapidly rotating black hole, $a=-0.9M$.  The computational domain covers only the northern hemisphere, $x\in[0,1]$, on which we solve for the even and odd parts of the field (Sec.~\ref{sec:mdsm}). For presentation we recombine them through Eq.~\eqref{eq:master_Func_even_odd} and plot over both hemispheres, $x\in[-1,1]$, which makes the lack of reflection symmetry across the equatorial plane explicit. 

Figure~\ref{fig:a0_r10_m2_N32} shows both the residual field $\bar\psi^{\rm R}$ and the retarded field $\bar\psi^{\rm ret}$. They differ only in $\mathcal D_2$, where $\bar\psi^{\rm ret}=\bar\psi^{\rm R}+\bar\psi^{\rm P}$; the field computed in $\mathcal D_1$ is the retarded field. The residual field therefore jumps across the $\mathcal D_1$--$\mathcal D_2$ interface by the value of the puncture, as imposed by the jump conditions of Sec.~\ref{sec:BCs}. Once the puncture is added, the retarded field is smooth across all interfaces, a non-trivial check of the consistency between the puncture, the effective source and the jump conditions. The behaviour at the poles follows from Sec.~\ref{sec:BCs}. For $m=2$ only the south-pole condition is imposed ($\delta_1=4$, $\delta_2=0$), and $\bar\psi$ is non-zero at the north pole. For $m=10$, $\delta_1$ and $\delta_2$ are both positive and $\bar\psi$ vanishes (within precision) at both poles.

\subsection{Convergence}\label{sec:convergence}

\begin{figure*}
    \centering
    \includegraphics[width=0.9\linewidth]{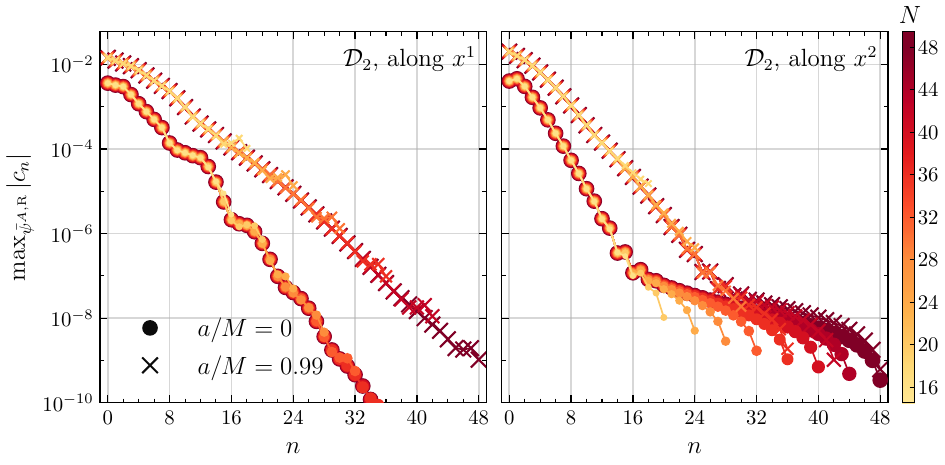} \\
    \caption{Convergence of the residual field in the particle domain $\mathcal D_2$, for $m=2$, $r_0/M=10$, and $a/M=0$ (dots) and $a/M=0.99$ (crosses). For each case, we show the Chebyshev coefficients of the slowest-converging of the four real fields $\bar\psi^{A,\rm R}$, for $N=16,\ldots,48$. Along $x^1$ the convergence is exponential. Along $x^2$, towards the particle, it becomes algebraic, limited by the order of the puncture. All other domains converge exponentially in both directions.}
    \label{fig: convergence a0_m2}
\end{figure*}

Figure~\ref{fig: convergence a0_m2} shows the convergence in the particle domain $\mathcal D_2$, where it is slowest. From the residual field solutions $\bar{\psi}_{m=2}^{A,\rm{R}}$, the worst convergence selected and the Chebyshev coefficients are plotted for $N=[16,\ldots,48]$ for both the Schwarzschild limit $a/M=0$ and an extremal-spin $a/M=0.99$ . Along $x^1$ the coefficients decay exponentially, down to $\sim10^{-14}$, for $a=0$ and down to $\sim 10^{-9}$ for $a=0.99M$. Along $x^2$, the direction towards the particle, initial exponential convergence gives way to algebraic convergence. The transition between the two behaviours varies with the choice of $a/M$ -- an effect related to the size of the worldtube for either spins (see Figure~\ref{fig: domain scaling} for an illustration of this effect). An algebraic rate of convergence is expected: the smoothness of the residual field at the particle is limited by the order of the puncture (Sec.~\ref{sec:puncture}), and the grid along $x^2$ extends to the particle. In the other domains, which do not contain the particle, the convergence is exponential in both directions with sufficient resolution. The algebraic error in $\mathcal D_2$ enters the other domains through the interface conditions and sets the accuracy of the fluxes (Sec.~\ref{sec:fluxes} and Appendix~\ref{app:worldtube}). Higher-order punctures, or local refinement around the particle, would improve the convergence rate, but are beyond the scope of this work.

\begin{figure}
    \centering
    \includegraphics[width=\linewidth]{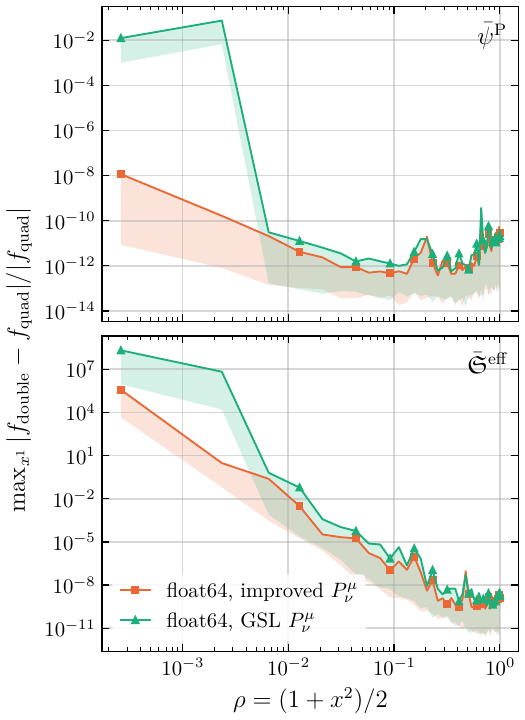}
    \caption{Accuracy of the puncture $\bar\psi^{\rm P}$ and of the effective source $\bar{\frak S}^{\rm eff}$ evaluated in double precision, relative to extended precision, as a function of the distance to the particle, $\rho=(1+x^2)/2$. We compare the default evaluation with the \texttt{GSL} associated Legendre functions (constructed from \texttt{gsl\_sf\_hyperg\_2F1\_renorm}) and the improved evaluation from Ref.~\cite{Vu:2026ypc} near the particle. Solid lines show the worst agreement, and shaded areas the range over $a/M\in\{\pm0.9,\pm0.99,\pm0.999,\pm0.9999\}$ and $m=1,2,3$, at fixed $r_0/M=10$. Close to the particle, double precision loses all accuracy in $\bar{\frak S}^{\rm eff}$, which is why both quantities are evaluated in quadruple precision.}
    \label{fig: puncture precision}
\end{figure}

To reach this level of convergence, the puncture and effective source must be evaluated in extended (quadruple) precision. Figure~\ref{fig: puncture precision} compares them with two double-precision evaluations for $m=1,2,3$ and near-extremal spins $|a|/M=0.9,\ldots,0.9999$: one evaluation using the associated Legendre functions constructed from the regularised hypergeometric functions in \texttt{GSL} (\texttt{gsl\_sf\_hyperg\_2F1\_renorm}) and one using an improved implementation, presented in Sec.~III~B of Ref.~\cite{Vu:2026ypc}, that incorporates a direct evaluation of a series representation of the hypergeometric function in areas of the parameter space where \texttt{GSL} fails to give an accurate value. For distances from the particle $\rho = (1+x^2)/2 \sim 10^{-2}$, corresponding to $N \sim 20$, the improved calculation for $\bar{\frak S}^{\rm eff}$ at double precision has a relative error of $\mathcal{O}(1)$ with respect to the extended precision values. The effect on $\bar \psi^{\rm P}$ is milder and can mostly be mitigated by using a better approximation for $P^\mu_\nu$ near the particle.

\subsection{The static mode \texorpdfstring{$m=0$}{m=0}}\label{sec:m0}

The mode $m=0$ is static, $\omega_m=0$, and does not radiate: $\dot E^\infty_0=\dot E^{\rm H}_0=0$. It is nonetheless part of the self-force, and in $m$-mode and frequency-domain calculations it typically requires separate treatment. In the Lorenz gauge, the low multipoles of the static mode contain gauge modes and the perturbations of the mass and angular momentum of the black hole, which need special treatment~\cite{Berndtson:2007gsc,Dolan:2012jg,Dolan:2023enf,lmtompaper}. In time-domain $m$-mode evolutions, the $m=0$ and $m=1$ modes suffer from gauge instabilities~\cite{Dolan:2012jg} (although see Ref.~\cite{Thornburg:2026eeb} for recent progress at alleviating this problem in $\ell m$-modes).  In the frequency domain, the static mode likewise requires special consideration, both in the elliptic $m$-mode approach of Ref.~\cite{Osburn:2022bby} and in approaches based on radial ODEs, where the boundary conditions at the horizon and at infinity are based on $e^{\pm i\omega r_*}$ and degenerate for $\omega\to0$, so that separate homogeneous solutions are needed~\cite{Akcay:2010dx}.

In our formulation the static mode requires no special treatment: the code solves it with the same operator, grids and boundary conditions as every other mode. Two properties make this possible. First, $\psi_4$ is gauge invariant and its multipoles start at $\ell=2$, so neither gauge modes nor the $\ell=0,1$ perturbations appear. Second, no external boundary data are imposed at the horizon or at $\scri^+$; the physical solution is selected by regularity alone, and this selection remains valid at $\omega_m=0$. The unphysical static branches behave as $(1-\sigma)^{\spin}$ at the horizon and as $\sigma^{1-\ell}$ at $\scri^+$, and both are singular for $\spin=-2$, $\ell\ge2$ (Sec.~\ref{sec:radial_reg}).

\begin{figure}
    \centering
   \includegraphics[width=\linewidth]{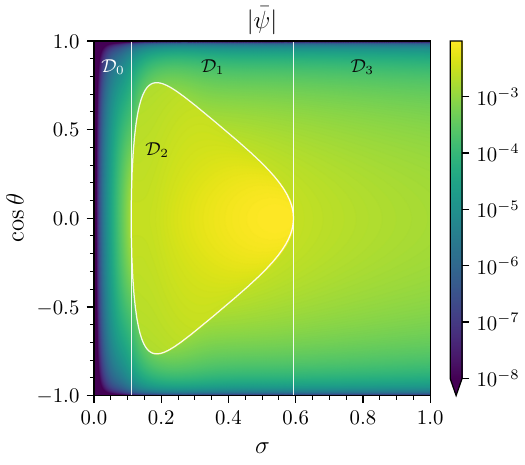}\\ \includegraphics[width=\linewidth]{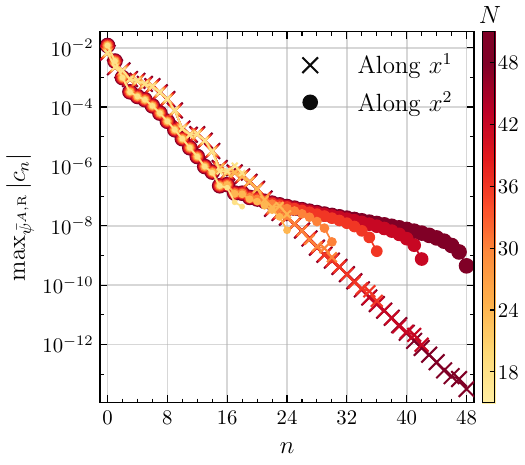}
    \caption{\emph{Top:} Absolute value $\bar\psi^{\text{ret}}$ for in unsourced domains and $\bar\psi^R$ in $\mathcal D_2$ for $a/M = 0.5$, $r_0/M=10$, $m=0$. \emph{Bottom:} Worst convergence of $\bar\psi^{A, \rm R}$ in $\mathcal{D}_2$ for the same solution. Convergence in $x^2$ is again algebraic while $x^1$ is exponential.} 
    \label{fig: convergence a0.5_m0}
    \end{figure}

Figure~\ref{fig: convergence a0.5_m0} shows the $m=0$ field for $a=0.5M$ and its convergence, which does not differ from that of the radiative modes. Since the fluxes vanish, we validate the static mode through its mode amplitudes. For $a=0$ we project the solution at fixed radii outside ($\sigma=0.02$, $0.06$, $0.1$) and inside ($\sigma=0.65$, $0.8$, $0.95$) the orbit onto spin-weighted spherical harmonics. We then compare the complex amplitudes $\ell$ by $\ell$, for $\ell=2,\ldots,5$, with those of the BHPToolkit, after accounting for the factor $Z$ relating $\bar\psi$ to the Boyer--Lindquist $\Psi_{-2}$. The relative difference of the dominant ($\ell = 2$) amplitude of the $m = 0$ mode decreases from $10^{-4}$ at $N = 16$ to $10^{-8}$ at $N = 32$ (Table~\ref{tab:amplitudes_a0}), and is the same at every radius on a given side of the orbit. This is consistent with the level of relative differences in the radiative modes $m = 1, 2, 3$ at these resolutions.

\begin{table}[h]
\footnotesize
\setlength{\tabcolsep}{2pt}
\renewcommand{\arraystretch}{1.2}
\begin{tabular}{r|c|c|c|c}
\hline\hline
 & \multicolumn{4}{c}{$m$} \\ \hline
$N$ & $0$ & $1$ & $2$ & $3$ \\
\hline
$16$ & $5.0\times10^{-4}$ & $3.0\times10^{-3}$ & $1.0\times10^{-3}$ & $9.9\times10^{-4}$ \\
$20$ & $1.3\times10^{-4}$ & $8.9\times10^{-5}$ & $1.1\times10^{-4}$ & $1.1\times10^{-4}$ \\
$24$ & $4.9\times10^{-6}$ & $1.6\times10^{-5}$ & $1.1\times10^{-5}$ & $7.9\times10^{-6}$ \\
$28$ & $5.9\times10^{-7}$ & $2.3\times10^{-6}$ & $6.0\times10^{-7}$ & $6.1\times10^{-7}$ \\
$32$ & $5.8\times10^{-8}$ & $1.3\times10^{-7}$ & $5.2\times10^{-8}$ & $4.7\times10^{-8}$ \\
\hline\hline
\end{tabular}
\caption{Relative difference $|R_{\ell m}-R_{\ell m}^{\rm BHPT}|/|R_{\ell m}^{\rm BHPT}|$ between the complex radial amplitudes of the $a=0$, $r_0=10M$ solution, projected onto spin-weighted spherical harmonics at $\sigma=0.02, 0.06, 0.1$ (outside the orbit) and $\sigma=0.65, 0.8, 0.95$ (inside), and those of the BHPToolkit, for the dominant multipole $\ell=\max(2,|m|)$; each entry is the largest over the radii. All multipoles $\ell\le5$ agree to within $10^{-6}$ at $N=32$.}
\label{tab:amplitudes_a0}
\end{table}

\subsection{Energy fluxes}\label{sec:fluxes}

The energy flux to infinity follows from the field at $\scri^+$. With our $m$-mode normalisation,
\be
\dot E^\infty_m = \dfrac{1}{8\pi^2\omega_m^2}\int_{-1}^{1}\lim_{r\to\infty}\big|r\,\psi_{4,m}\big|^2\,dx
= \dfrac{1}{(2\pi)^2}\sum_\ell \dfrac{|c_{\ell m}|^2}{4\pi\omega_m^2},
\label{eq:flux_inf}
\ee
where $c_{\ell m}$ are the coefficients of $\lim r\psi_{4,m}$ in the basis of spin-weighted spheroidal harmonics $S_{\ell m}(x;a\omega_m)$ normalised as $2\pi\int S^2_{\ell m}dx=1$. We evaluate the right-hand side by projecting the numerical solution on $\sigma=0$ onto $S_{\ell m}$ ($\ell\le20$). This is preferable to integrating $|r\psi_{4,m}|^2$ directly: for large $|m|$, the factors $(1\pm x)^{-\delta_{1,2}/2}$ of Eq.~\eqref{eq:reg_axis_ansatz} amplify round-off near the poles.

The flux through the horizon follows from the field at $\sigma=1$. Near the horizon, each $\ell$-mode of $\Psi_{-2}=\zeta^4\psi_4$ behaves as $Z^H_{\ell m}\,\Delta^2 e^{-ik r_*}$, with $k=\omega_m - m\Omega_{\rm h}$, and the regularised field at $\sigma=1$ directly provides $|Z^H_{\ell m}|$ after projection onto $S_{\ell m}$. The Teukolsky--Press formula then gives
\bea
\dot E^{\rm H}_m &=& \sum_\ell \dfrac{\alpha_{\ell m}\,|Z^H_{\ell m}|^2}{4\pi\omega_m^2},\\
\alpha_{\ell m} &=& \dfrac{256\,(2M\rh)^5\,k\,(k^2+4\epsilon^2)(k^2+16\epsilon^2)\,\omega_m^3}{|C_{\ell m}|^2},\hspace{12pt}
\eea
with $\epsilon=\sqrt{M^2-a^2}/(4M\rh)$ and
\begin{align}
    |C_{\ell m}|^2 ={}& \big[(\lambda_{\ell m}+2)^2 + 4a\omega m - 4a^2\omega^2\big] \nn
        & \times\big(\lambda_{\ell m}^2+36a\omega m - 36a^2\omega^2\big) \nn
        & + (2\lambda_{\ell m}+3)\big(96a^2\omega^2 - 48 a\omega m\big) \nn
        & + 144\,\omega^2(M^2-a^2),
\end{align}
where $\lambda_{\ell m}$ is the spheroidal eigenvalue and $\omega=\omega_m$. We have checked that this formula, with the BHPToolkit normalisation $R^{\rm in}\to\Delta^2e^{-ikr_*}$, reproduces the BHPToolkit horizon fluxes to $10^{-12}$.

\begin{figure*}[t!]
	\centering
	\includegraphics[width=\textwidth]{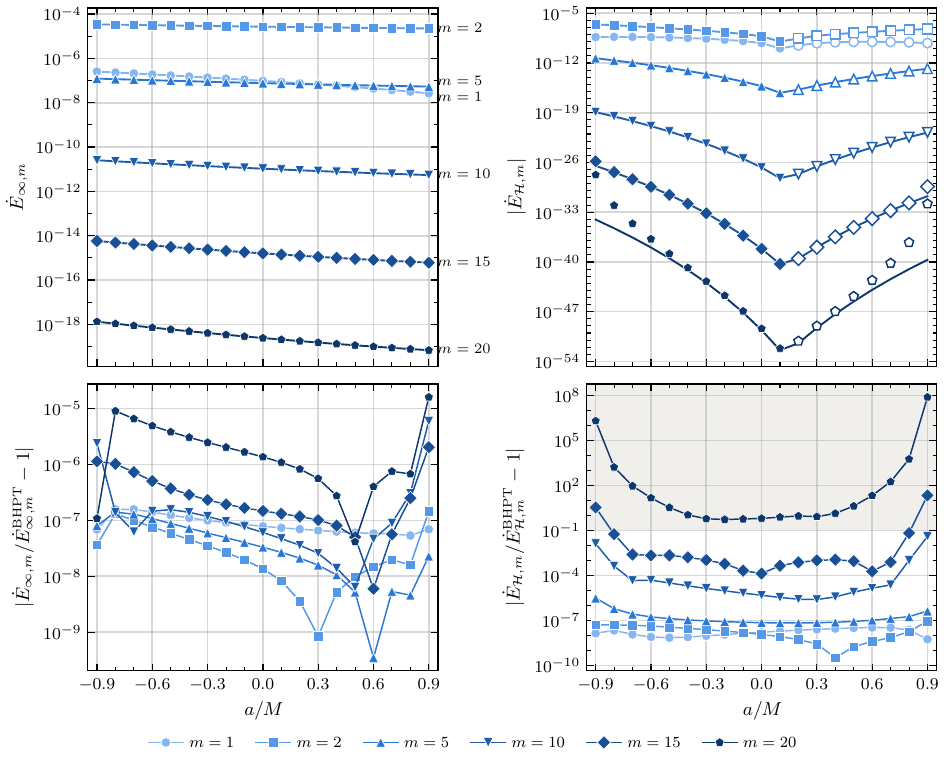}
	\caption{Comparison of energy fluxes at \scri~ and $\mathcal{H}$~for $r_0=10M$ against the BHPToolkit. \emph{Top:} markers denote calculated fluxes while Toolkit values are denoted by solid lines. Colours and marker types denote $m$-modes. Sampled in steps of $\Delta a = 0.1$ from $a=[-0.9,0.9]$. Open marker represent superradiant fluxes. \emph{Bottom:} Relative difference to BHPToolkit values of per-$m$ energy fluxes. The shaded region in the horizon flux plot denotes non-convergent results. Modes $m\le10$ use $N=40$, modes $m\ge11$ use $N=48$.}
	\label{fig:fluxes}
\end{figure*}

Table~\ref{tab:flux_r10} and Fig.~\ref{fig:fluxes} summarise the comparison with \texttt{pybhpt}. For $m\le5$ and all five spins, the fluxes at infinity agree to $\lesssim1.4\times10^{-7}$, and the horizon fluxes, although much smaller, to $\lesssim3\times10^{-6}$. For fixed $N$, the agreement degrades slowly as $m$ and $|a|$ increase, the largest differences at infinity, $6\times10^{-6}$ and $1.6\times10^{-5}$, occurring for $a=0.9M$ and $m=10$ and $20$. Since the flux is dominated by the lowest modes, the flux summed over $m=1$--$20$ agrees to $\lesssim1.3\times10^{-7}$ for all five spins. Figure~\ref{fig:fluxes} shows the same level of agreement across $a/M\in[-0.9,0.9]$, including the superradiant modes, whose horizon flux is negative.

The horizon fluxes of the highest modes are poorly resolved at these resolutions. For $|a|=0.9M$, the relative difference reaches $10^{-2}$ at $m=10$, and the modes $m\ge15$ are not resolved. For $|a|\le0.5M$, only $m=20$ is not resolved (dashes in Table~\ref{tab:flux_r10}; shaded region in Fig.~\ref{fig:fluxes}). These horizon fluxes are smaller than the corresponding fluxes at infinity by many orders of magnitude, so they do not affect the total flux or the dissipative self-force. They can be recovered by shrinking the worldtube size (Appendix~\ref{app:worldtube}) and increasing the resolution $N$.

The accuracy at fixed $N$ also depends on the size of the worldtube. With the worldtube scaled according to~\ref{e: worldtube scaling}, the worst error in $\dot E_\infty$ over $m \in \{1, 10, 15, 20\}$ and $|a| \leq 0.9M$ drops from $1.6\times10^{-5}$ to $2\times10^{-8}$ at $N = 48$, and the worst $\dot E_\mathcal{H}$ error over $m \in \{1, 10\}$ drops from $1.5\times10^{-3}$ to $7\times10^{-9}$. The remaining error is then set by the algebraic convergence along $x^2$ in $\mathcal{D}_2$: for $N \geq 40$ the flux errors decrease approximately as $N^{-4}$ (see Appendix~\ref{app:worldtube} for a further discussion on the dependence of worldtube sizes on $a$ and error mitigation through worldtube rescaling).

\begin{table*}[t]
\centering
\footnotesize
\setlength{\tabcolsep}{2pt}
\renewcommand{\arraystretch}{1.2}
\begin{tabular}{rc|ccccccc|c}
\hline\hline
 & & \multicolumn{7}{c|}{$m$} & \\
$a/M$  &               &          1          &          2          &          3          &          5          &          10         &          15         &          20         &              $\sum$\\
\hline
$0$    & $\Delta\dot{E}_\infty$      & $-7.8\times10^{-8}$ & $-1.4\times10^{-8}$ & $-2.0\times10^{-8}$ & $-3.3\times10^{-8}$ & $-6.1\times10^{-8}$ & $+1.5\times10^{-7}$ & $+1.4\times10^{-6}$ &  {\boldmath$-1.5\times10^{-8}$}\\
       & $\Delta\dot{E}_\mathcal{H}$ & $-1.7\times10^{-8}$ & $-1.2\times10^{-8}$ & $-1.9\times10^{-8}$ & $-7.1\times10^{-8}$ & $-4.7\times10^{-6}$ & $+1.3\times10^{-4}$ &          --         &  {\boldmath$-1.2\times10^{-8}$}\\
\hline
$0.5$  & $\Delta\dot{E}_\infty$      & $-6.0\times10^{-8}$ & $+9.8\times10^{-9}$ & $+1.5\times10^{-9}$ & $-5.2\times10^{-9}$ & $+6.4\times10^{-9}$ & $+5.0\times10^{-8}$ & $-4.1\times10^{-8}$ &  {\boldmath$+8.6\times10^{-9}$}\\
       & $\Delta\dot{E}_\mathcal{H}$ & $-3.1\times10^{-8}$ & $+1.8\times10^{-9}$ & $-1.2\times10^{-8}$ & $-8.6\times10^{-8}$ & $-7.7\times10^{-6}$ & $+8.9\times10^{-4}$ &          --         & {\boldmath$-1.5\times10^{-11}$}\\
\hline
$0.9$  & $\Delta\dot{E}_\infty$      & $+6.9\times10^{-8}$ & $-1.4\times10^{-7}$ & $-4.4\times10^{-8}$ & $-2.3\times10^{-8}$ & $-6.0\times10^{-6}$ & $-2.0\times10^{-6}$ & $-1.6\times10^{-5}$ &  {\boldmath$-1.3\times10^{-7}$}\\
       & $\Delta\dot{E}_\mathcal{H}$ & $+5.6\times10^{-9}$ & $+9.1\times10^{-8}$ & $+1.2\times10^{-7}$ & $+4.2\times10^{-7}$ & $+4.0\times10^{-2}$ &          --         &          --         &  {\boldmath$+9.1\times10^{-8}$}\\
\hline
$-0.5$ & $\Delta\dot{E}_\infty$      & $-1.2\times10^{-7}$ & $-5.9\times10^{-8}$ & $-6.6\times10^{-8}$ & $-8.8\times10^{-8}$ & $-1.6\times10^{-7}$ & $+3.7\times10^{-7}$ & $+3.8\times10^{-6}$ &  {\boldmath$-6.0\times10^{-8}$}\\
       & $\Delta\dot{E}_\mathcal{H}$ & $-7.1\times10^{-9}$ & $-3.6\times10^{-8}$ & $-4.2\times10^{-8}$ & $-1.3\times10^{-7}$ & $-3.1\times10^{-5}$ & $-2.2\times10^{-3}$ &          --         &  {\boldmath$-3.4\times10^{-8}$}\\
\hline
$-0.9$ & $\Delta\dot{E}_\infty$      & $+6.9\times10^{-8}$ & $-3.7\times10^{-8}$ & $+9.2\times10^{-9}$ & $-8.1\times10^{-8}$ & $-2.4\times10^{-6}$ & $-1.1\times10^{-6}$ & $-1.1\times10^{-7}$ &  {\boldmath$-3.0\times10^{-8}$}\\
       & $\Delta\dot{E}_\mathcal{H}$ & $-1.4\times10^{-8}$ & $-5.0\times10^{-8}$ & $-2.3\times10^{-7}$ & $-3.1\times10^{-6}$ & $+1.3\times10^{-2}$ &          --         &          --         &  {\boldmath$-5.4\times10^{-8}$}\\
\hline\hline
\end{tabular}
\caption{Relative difference $\Delta \dot E =\dot E/\dot E_{\mathrm{BHPT}}-1$ between the per-$m$ energy fluxes at \scri~and at the horizon and the reference values from \texttt{pybhpt}, for circular equatorial orbits at $r_0=10M$. Modes $m\le10$ use $N=40$, modes $m\ge11$ use $N=48$. The $\sum$ column gives the relative difference of the fluxes summed over $m=1$--$20$. Dashes mark horizon fluxes that are not resolved due to high relative error (smallest absolute values).}
\label{tab:flux_r10}
\end{table*}

\section{Conclusion}\label{sec:conclusion}

We have presented a hyperboloidal, multi-domain spectral solver for the $m$-mode spin-weight $-2$ Teukolsky equation with a point-particle source on circular equatorial orbits in Kerr spacetime, based on the puncture/effective-source scheme. The breaking of equatorial symmetry by the spin weight and by the black hole spin is handled by an even/odd decomposition under $x\to-x$. Besides the regularity conditions at the horizon, future null infinity and the equator, which are imposed through the field equations, the spin-weighted problem requires Dirichlet conditions at the poles, without with the linear system is ill-posed.

With these ingredients, for $r_0=10M$ and $a/M=0,\pm0.5,\pm0.9$, prograde and retrograde orbits alike, the energy fluxes to infinity agree with the BHPToolkit to $\lesssim1.4\times10^{-7}$ for the dominant modes $m\le5$ and to better than $2\times10^{-5}$ for all $m\le20$, and the horizon fluxes to $\lesssim3\times10^{-6}$ for $m\le5$. We obtain the dissipative self-force from the flux balance, with the sum over $m$ accurate to $\lesssim1.3\times10^{-7}$. The static mode $m=0$ is solved in exactly the same way as the radiative modes. A single mode takes about $75\,$s at $N=40$ on a laptop, dominated at high resolution by the sparse $LU$ factorisation (Appendix~\ref{app: performance}), and the modes are independent and can be computed in parallel. The accuracy is currently limited by the order of the puncture, through the algebraic convergence in the particle domain. 

The numerical method carries over to second order. In the Teukolsky formulation of second-order self-force, the second-order Weyl scalar satisfies the same Teukolsky operator as at first order, with a source that is quadratic in the first-order metric perturbation~\cite{Campanelli:1998jv,Spiers:2023cip}. The operator, the grids, the boundary treatment at the horizon, null infinity and the poles, and the sparse direct solver of this paper can therefore be reused unchanged; only the source changes. The solver takes the effective source as data on the two-dimensional grid of each domain, so it can be applied to any $m$-mode source, including a second-order source computed elsewhere in the second-order programme. Two features of the method suit the second-order problem. First, the $m$-mode decomposition avoids the coupling between $(\ell,m)$-modes: since the background is axisymmetric, each $m$-mode of a quadratic source is a single sum over products of first-order $m$-modes, which our solver provides directly on the two-dimensional grid. Second, the second-order source does not decay with compact support, but extends to the horizon and to infinity; the hyperboloidal slices include both boundaries in the numerical domain, so no truncation or matching of the source is needed. The particle domain, adapted to the comoving coordinates of the particle, is where the stronger singularity of the second-order source must be resolved, and this is where higher-order punctures, local refinement, or a subdivision of the particle domain will be most important.

This work complements other first-order Kerr infrastructures that are being developed with second order in mind. These include the frequency-domain Lorenz-gauge calculations for circular equatorial orbits~\cite{Dolan:2021ijg,Dolan:2023enf}, the elliptic $m$-mode approach~\cite{Osburn:2022bby}, and the GHZ-Teukolsky puncture scheme~\cite{Bourg:2024vre}. In addition, the contemporaneous work of Ref.~\cite{Nishimura2026} computes the $m$-modes of the Lorenz-gauge metric perturbation in Kerr directly, using the elliptic solver of the \texttt{SpECTRE} code~\cite{spectrecode}, so that the metric perturbation is obtained without reconstruction. It will be interesting to compare the two approaches: since $\psi_4$ is gauge invariant, it can be computed from their Lorenz-gauge $m$-modes and compared with our solution mode by mode.

Further extensions are higher-order punctures, which would remove the present limit on the accuracy from the algebraic convergence in the particle domain, and more general orbits. For eccentric equatorial orbits the source contains a discrete set of frequencies $\omega_{mn}$, and each $(m,n)$-mode is solved with the same operator. For inclined orbits the particle is no longer at a fixed polar angle, and the even/odd decomposition and the particle domain must be adapted. Finally, the infrastructure is not tied to the Teukolsky equation: we have obtained preliminary results for the $m$-modes of the Lorenz-gauge metric perturbation in Kerr with the same multi-domain solver, which will be presented in future work.


\begin{acknowledgments}
The authors gratefully acknowledge Adam Pound for helpful discussions.

R.P.M. is thankful for the hospitality of the UCD Relativity Group during several visits while the project was developed. This project initiated thanks to the support of EuCAPT International Exchange Award. He acknowledges support from the Villum Investigator program supported by the VILLUM Foundation (grant no.\ VIL37766) and the DNRF Chair program (grant no.\ DNRF162) by the Danish National Research Foundation. The Center of Gravity is a Center of Excellence funded by the Danish National Research Foundation under grant No. DNRF184. 

 C.P.\ acknowledges support from a Royal Society -- Research Ireland University Research Fellowship via grant URF/R1/211027.

SDU acknowledges the support of a Royal Society University Research Fellowship and the ERC Consolidator/UKRI Frontier Research Grant GWModels (selected by the ERC and funded by UKRI [grant number EP/Y008251/1]).
This work is supported by ERC grant EMRIWaveforms (DOI: \href{https://doi.org/10.3030/101200625}{10.3030/101200625}).

This work makes use of the Black Hole Perturbation Toolkit~\cite{BHPToolkit}, Wolfram Mathematica~\cite{Mathematica} and xAct~\cite{xAct,Martin-Garcia:2008ysv,xCoba}.

\paragraph*{Statement on AI Use.}The authors acknowledge the use of generative AI tools to assist in coding and text editing. All AI-generated suggestions were reviewed, revised, and approved by the authors, who take full responsibility for the accuracy and integrity of the work.
\end{acknowledgments}


\appendix


\section{Performance Metrics}
\label{app: performance}

While the spectral basis of the solving method provides highly accurate solutions, it comes at the cost of dense matrix representations in each domain. The overall density of the discrete Jacobian matrix across the four domains and four fields does not have the same extreme level of sparsity as is present for finite difference schemes. This means that for higher-resolution solutions, memory usage and timings need to be considered. Here we include performance metrics such as individual module and overall walltimes with spectral resolution across different parts of the $(a,m,r_0)$ phase space. For all data below, runs were performed using Intel(R) Core(TM) i7-9750H CPU at 2.60GHz with 32.0 GB of RAM and 12 OMP threads. Peak memory usage during runs scales roughly as $(N+1)^4$, with resolutions of $N=16$ requiring 0.1 GB (averaged over 12 runs) while resolutions of $N=48$ require 6.0 GB (averaged over 200+ runs).

\subsection{Timings}\label{sec:timing}

A single $m$-mode solve consists of three main stages: the evaluation of the puncture and effective source, in quadruple precision, on the $(N+1)^2$ Chebyshev grid of $\mathcal D_2$ (Sec.~\ref{sec:puncture}); the assembly of $\hat J$; and the solution of the linear system by sparse $LU$ factorisation followed by iterative refinement. The first two stages utilise 12 OpenMP threads while the linear solver uses a single thread. The number of unknowns is $n_{\rm total}=16(N+1)^2$, from $10\,000$ at $N=24$ to $38\,416$ at $N=48$, and the number of non-zero entries of $\hat J$ grows as $\sim N^{3.8}$, from $3.6\times10^6$ to $5.0\times10^7$: each row couples to about half of the $(N+1)^2$ points of its domain.

\begin{figure*}
    \centering
    \includegraphics[width=0.75\textwidth]{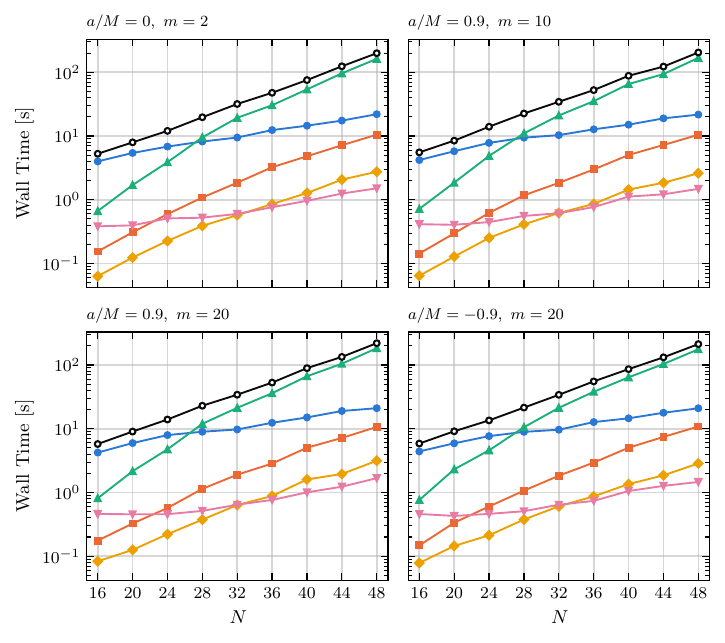}
    \caption{Wall time of each stage of a single solve as a function of $N$, for different black hole spins and $m$-modes: linear solve and iterative refinement (yellow diamonds), assembly of the Jacobian $\hat J$ (orange squares), grid initialisation and I/O (pink downward triangles), $LU$ factorisation on a single thread (green upward triangles), construction of $\bar\psi^{\rm P}$ and $\bar{\frak S}^{\rm eff}$ (blue filled circles), and total (black empty circles). The four cases are almost indistinguishable: the cost does not depend on $a$ or $m$. The sparse $LU$ factorisation grows fastest, $\propto N^{5}$, and dominates the total cost for $N\gtrsim 28$.}
    \label{fig: walltime scaling}
\end{figure*}

\begin{figure}
    \centering
    \includegraphics[width=\linewidth]{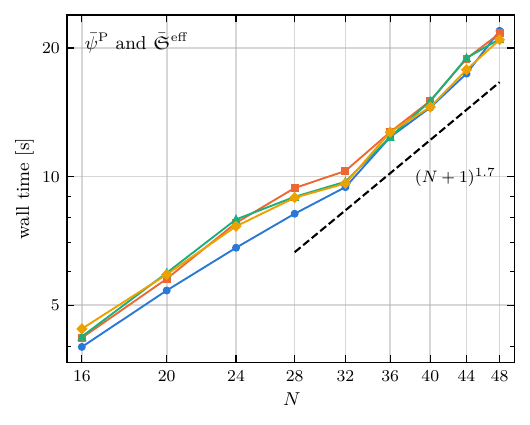}
    \caption{Wall time of the evaluation of the puncture and effective source in quadruple precision, for $a/M=0$, $m=2$ (blue circles), $a/M=0.9$, $m=10$ (orange squares), $a/M=0.9$, $m=20$ (green triangles) and $a/M=-0.9$, $m=20$ (yellow diamonds). Since the evaluation is done point by point, in parallel over the grid, its cost grows only as $\sim(N+1)^{1.7}$, and it does not depend on $a$ or $m$.}
    \label{fig: seff timings}
\end{figure}

Figure~\ref{fig: walltime scaling} breaks the wall time of a single solve down into its modules for four combinations of $a$ and $m$. Fitted for $N\ge28$, the total time grows as $N^{4.1}$, but the individual modules scale very differently. The puncture and effective source are evaluated point by point in quadruple precision, in parallel over the grid points. Their cost grows only as $\sim(N+1)^{1.7}$, from $4\,$s at $N=16$ to $21\,$s at $N=48$, and is the same for all $a$ and $m$ considered (Fig.~\ref{fig: seff timings}). The assembly time of $\hat J$ grows as $N^{4.1}$, close to its number of non-zero entries ($\sim N^{3.8}$), but stays below the cost of the puncture up to $N=48$ ($10\,$s against $21\,$s). The sparse $LU$ factorisation, on a single thread, grows fastest, as $N^{5.0}$. It overtakes the puncture evaluation at $N\approx28$ and accounts for over $80\%$ of the total at $N=48$. The subsequent solve with iterative refinement ($\propto N^{3.5}$) takes at most a few seconds, and the grid set-up and output take less than $2\,$s. The four cases of Fig.~\ref{fig: walltime scaling} are almost indistinguishable: at $N=48$ the total is $200$--$220\,$s in all of them, so the cost does not depend on $a$ or $m$. At the resolutions needed here, the cost is thus set by the factorisation, which is the natural target for further speed-up, e.g.\ with a multithreaded, Krylov iterative methods or GPU sparse direct solver.

\subsection{Matrix Conditioning}

\begin{figure*}
    \centering
    \includegraphics[width=\textwidth]{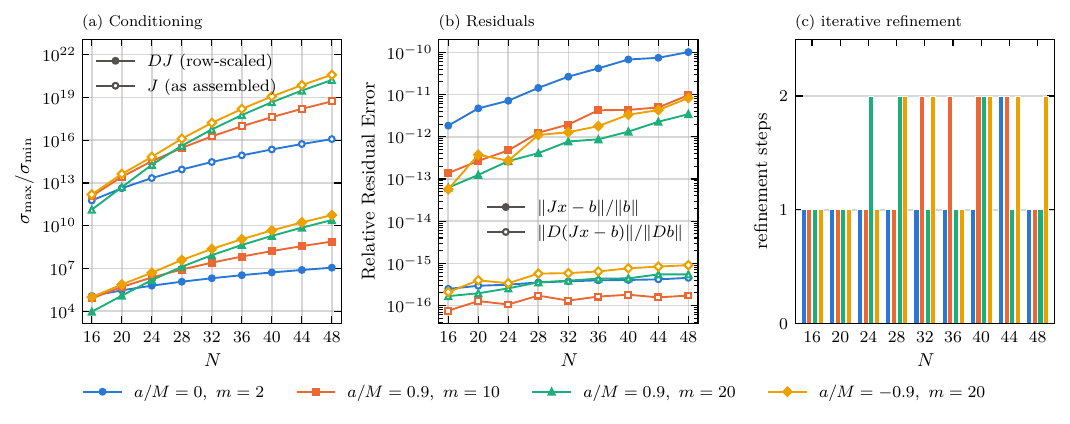}
   \caption{Measures of the discrete Jacobian $\hat J$ as a function of the spectral resolution $N$, for the four $(a,m)$ combinations shown, with $r_0/M=10$. (a) Condition number, i.e.\ the ratio of the largest and smallest singular values, of the matrix as assembled ($\hat J$, open symbols) and after row scaling ($D\hat J$, filled symbols). (b) Relative residual of the linear system after iterative refinement, unscaled ($\|\hat J\vec X-\vec S\|/\|\vec S\|$, filled symbols) and row-scaled ($\|D(\hat J\vec X-\vec S)\|/\|D\vec S\|$, open symbols). (c) Number of iterative-refinement steps, with the residual evaluated in quadruple precision, performed after the $LU$ factorisation to reach double-precision accuracy.}
    \label{fig:conditioning}
\end{figure*}

Figure~\ref{fig:conditioning} characterises the linear system~\eqref{eq:linear_system} for the same four combinations of $a$ and $m$. As assembled (according to Eq.~\eqref{eq:indexing_func}), $\hat J$ is badly conditioned. The ratio of its largest to smallest singular value is already $\sim10^{12}$ at $N=16$, and at $N=48$ it reaches $10^{16}$ for $a=0$, $m=2$ and $10^{21}$ for $a=-0.9M$, $m=20$. Most of this reflects the very different scales of the rows of $\hat J$, which contain the field equations in the four domains, the interface conditions and the pole conditions. After row scaling, $D\hat J$ with $D$ diagonal, the condition number is lower by $7$--$10$ orders of magnitude. It still grows with $N$, faster for larger $m$ and $|a|$, but stays between $\sim10^{7}$ ($a=0$, $m=2$) and $\sim10^{11}$ ($a=-0.9M$, $m=20$) at $N=48$.

The linear system is solved with a sparse $LU$ factorisation in double precision, followed by iterative refinement with the residual evaluated in quadruple precision. One or two refinement steps suffice at all resolutions (Fig.~\ref{fig:conditioning}c). The relative residual of the row-scaled system, $\|D(\hat J\vec X-\vec S)\|/\|D\vec S\|$, is then $10^{-16}$--$10^{-15}$, at the level of round-off, for all cases and resolutions. The unscaled residual, $\|\hat J\vec X-\vec S\|/\|\vec S\|$, is larger, $10^{-13}$--$10^{-10}$, because it is dominated by the rows with the largest entries. The usual bound on the relative error of the solution, the condition number of $D\hat J$ times the scaled residual, is $\lesssim10^{-9}$ for $a=0$, $m=2$ at $N=48$, well below the discretisation error. For $m=20$ at $|a|=0.9M$ it is $\sim10^{-5}$. This bound is pessimistic, but it shows that round-off in the linear solve may contribute to the error of the highest modes at the largest $N$, where the flux differences in Table~\ref{tab:flux_r10} reach $10^{-5}$.

\subsection{Puncture and Effective Source}
\begin{figure*}
    \centering
    \includegraphics[width=\textwidth]{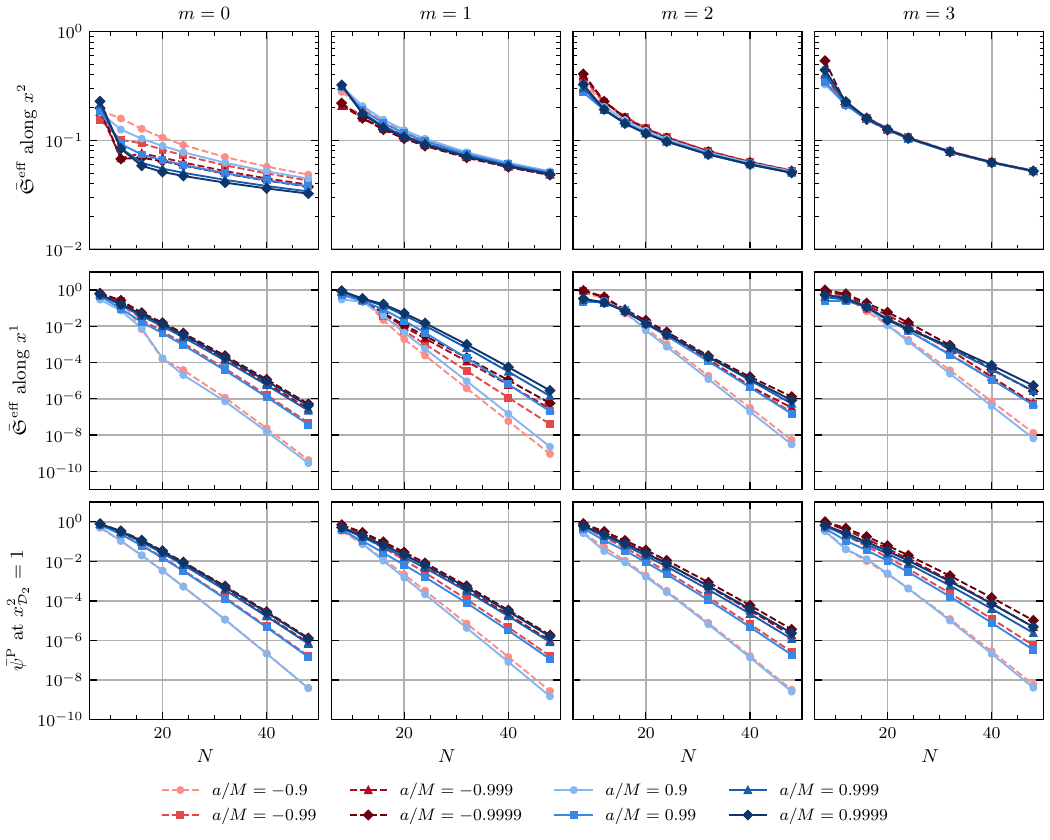}
    \caption{Convergence of the effective source $\bar{\frak S}^{\rm eff}$ and of the puncture on the $\mathcal D_1$--$\mathcal D_2$ interface, $\bar\psi^{\rm P}(x^2=1)$, as $|a|\to M$, for $m\in[0,3]$ with fixed $r_0=10M$. As a proxy for the truncation error, we show the relative size of the Chebyshev tail, i.e.\ the largest of the last three coefficients. Dashed blue (red) series correspond to prograde (retrograde) orbits with increasingly extremal spins. The convergence remains exponential as $|a|\to M$, except along $x^2$ for $\bar{\frak S}^{\rm eff}$, where it is algebraic, as expected.}
    \label{fig: alimit convergence}
\end{figure*}

To test that the puncture and effective source remain valid at near-extremal spins, we consider the convergence of the hyperboloidal effective source in both $x^1$ and $x^2$ directions as a function of $N$ for a set of near-extremal prograde and retrograde spins. Also relevant is the value of the puncture at the upper $\mathcal{D}_2$ boundary with $\mathcal{D}_1$, as this is used to calculate the jump conditions of Sec.~\ref{sec:BCs}. In Figure~\ref{fig: alimit convergence} we plot the normalized size of the Chebyshev coefficients of the tail (largest of $N-2,~N-1,~N$) as a proxy for truncation error. We can see that over $m=0,1,2,3$ the convergence of the effective source and puncture boundary value remains exponential where expected as $|a|\to1$. The exception is the convergence along $x^2$, which is algebraic as expected. 

\section{Domain Scaling \& Worldtube Size}
\label{app:worldtube}
Consider the scaling of the size of the worldtube with black hole spin. We have defined the bounds of the worldtube as
\begin{align}
    \sigma_\pm = \dfrac{\sigma_0}{1\mp \eta\,  \mathscr{B}\, \sigma_0}
\end{align}
where $\mathscr B$ is defined in Eq.~\eqref{eq:B_def}.
Thus, for unit mass, the size of the worldtube is a function of both the radial position of the particle as well as the spin of the primary. This is relevant when considering particle orbits near the innermost stable circle orbit (ISCO). In terms of the dimensionless spin $\chi = a/M$, the ISCO is
\begin{align}
    \dfrac{r_{\isco}}{M} &= 3 + Z_2 \mp \sqrt{(3 - Z_1)(3+Z_1+2Z_2)}
\end{align}
where $Z_1 = 1 + (1-\chi^2)^{1/3} \left((1+\chi)^{1/3} + (1-\chi)^{1/3} \right)$ and $Z_2 = \sqrt{3\chi^2 +Z_1^2}$. Because the ISCO of a near-extremal black hole lies at $r_{\isco} = 1.45M$ for prograde and $r_{\isco} = 8.97M$ for retrograde orbits ($|a| = 0.99M$), prograde orbits near the ISCO have compact worldtubes ($r_0/r_h \simeq 1.3$), whereas retrograde ones are roughly $r_0/r_h \gtrsim 7.9$. The convergence along $x^1$ in $\mathcal{D}_1$ and $\mathcal{D}_2$ remains exponential but slows as $r_0/r_h$ increases. Specifically, at $N = 32$ the relative error of the flux at \scri (worst of $|m| \leq 3$) is $8\times10^{-8}$ for $a = 0.99M,~r_0 = 1.5M$ but only $2 \times 10^{-4}$ for $|a| = 0.99M,~r_0 = 9M$, independent of the orbital direction. In both cases the attainable accuracy is ultimately limited by the algebraic (roughly $N^{-4}$) convergence along $x^2$ in $\mathcal{D}_2$, itself a function of the finite smoothness of the effective source at the particle.

\begin{figure*}
    \centering
    \includegraphics[width=\textwidth]{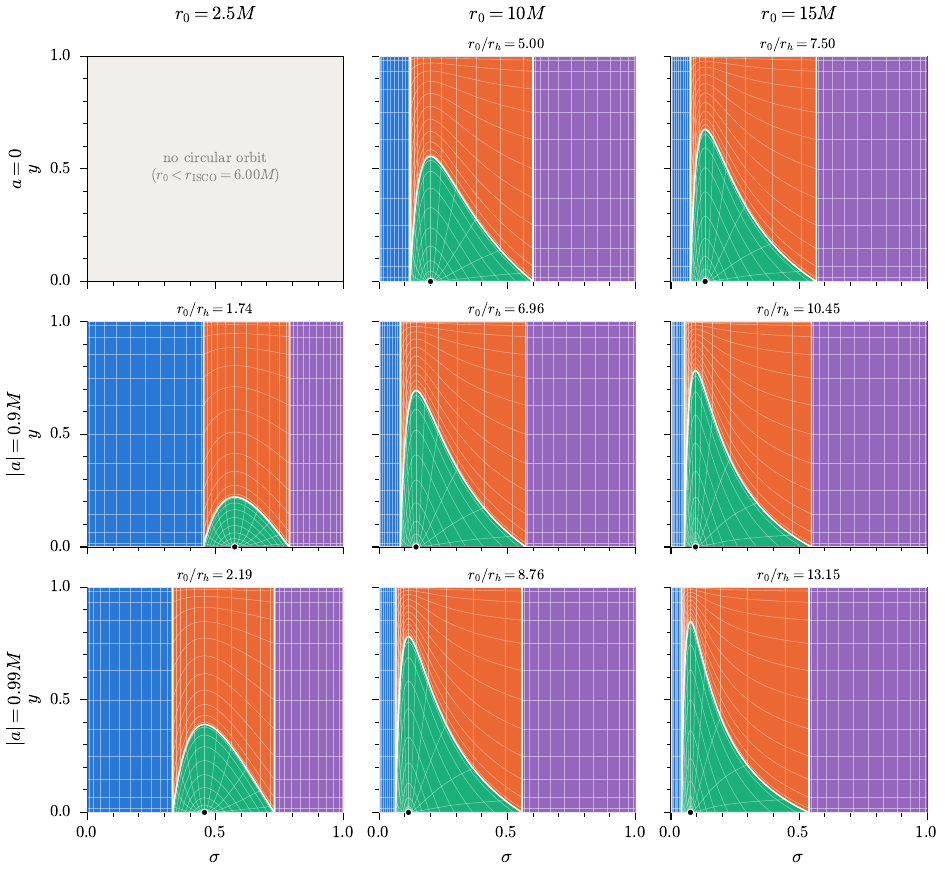}
    \caption{Scaling with spin and particle position for each domain: $\mathcal{D}_0$ (blue), $\mathcal D_1$ (orange), $\mathcal D_2$ (green), $\mathcal{D}_3$ (purple). The ratio of particle position to event horizon $r_0/r_h$ is shown above each panel.}
    \label{fig: domain scaling}
\end{figure*}

The rate of exponential convergence inside the worldtube is set by the worldtube's size. For $|a|=0.99M$, a prograde orbit at $r_0=1.5M$ reaches a relative \scri~flux error of $8\times10^{-8}$ at $N=32$, whereas orbits at $r_0=9M$ remain at $\sim1.5 \times 10^{-4}$ for prograde and $\sim3.8\times 10^{-4}$ for retrograde orbits. Because the prograde ISCO approaches the horizon as $a\to M$, strong-field prograde orbits are numerically easy; retrograde orbits, whose ISCO lies at large $r_0/r_h$, are more difficult because of the worldtube geometry.

\begin{figure*}
    \centering
    \includegraphics[width=\linewidth]{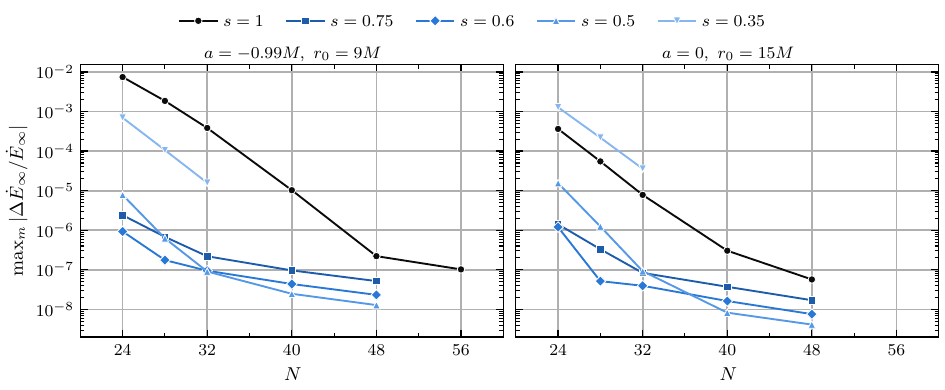}
    \caption{Effect of the worldtube scale $s$, with $\eta\to s\,\eta$. We show the largest relative error of the \scri~flux over $m=1,2$, against reference values from \texttt{pybhpt}, as a function of $N$, for a near-extremal retrograde orbit and for the Schwarzschild limit. Shrinking the worldtube improves the accuracy at fixed $N$, until the enlarged domains $\mathcal D_0$ and $\mathcal D_3$ limit the effective resolution.}
\label{fig:worldtube_scan}
\end{figure*}

To compensate, we introduce here a scale factor $0<s\le1$, which shrinks the worldtube by ${\eta \to s \eta}$, with $\mathcal{D}_0$ and $\mathcal{D}_3$ extended to remain attached (Fig.~\ref{fig:worldtube_scan}). At $N=32$, $s=0.5$ lowers the worst \scri~flux error over $m=1,2$ for $a=-0.99M$, $r_0=9M$ from $3.8\times10^{-4}$ to $9\times10^{-8}$, an accuracy that is comparable to $s=1$ at $N=56$ ($1.0\times 10^{-7}$). At $N=48$, $s=0.5$ gives relative flux errors of $1.3\times10^{-8}$ for this orbit and $4\times10^{-9}$ for $a=0$, $r_0=15M$, better than $s=1$ at $N=64$ ($5.7\times10^{-8}$ for $m = 1$) at about a quarter of the run time. However, the worldtube cannot be shrunk without limit: the sizes of $\mathcal{D}_0$ and $\mathcal{D}_3$, grow as the worldtube shrinks, reducing the effective resolution at fixed $N$ (Figure~\ref{fig: domain scaling}). By $s=0.35$, these truncation errors dominate. The best results lie at a reduced worldtube scale set by the empirical relation 
\begin{align}
    \label{e: worldtube scaling}
    s=\min\left(1,\dfrac{2(1+\sigma_0)}{5(1-\sigma_0)}\right).
\end{align}

Once the $x^1$ direction in $\mathcal{D}_2$ is resolved, the remaining error originates along $x^2$, i.e.\ away from the particle, where the finite smoothness of the effective source limits the Chebyshev expansion to algebraic convergence. Measured against a fixed normalisation, the tail of the Chebyshev series falls as $\,N^{-3.9}$ for every $s$ and in both prograde and retrograde orbits, and the flux errors at $N\ge40$ decrease at the same rate. Shrinking the worldtube thus lowers this floor only in proportion to $s$; the large gains are realized in the $x^1$ direction.

At $r_0=10M$, over $a=0$ to $|a|=0.9M$, the rule ~\eqref{e: worldtube scaling} reduces the worst \scri~flux (relative) error over $m\in\{1,10,15,20\}$ at $N=48$ from $1.6\times10^{-5}$ to $2\times10^{-8}$ against the \texttt{pybhpt} results, and it resolves the horizon flux for $|a|=0.9M$ up to $m=15$ ($3\times10^{-7}$ for the retrograde $m=15$ mode). The approach has two limitations. First, it leaves the algebraic floor in place; removing it requires local refinement around the particle (clustering of the $x^2$ grid, a subdivision of $\mathcal{D}_2$, or a discontinuous Galerkin discretisation there) or a higher-order puncture. Second, the enlarged $\mathcal{D}_3$ is forced to support a wider dynamic range. For $a=0$ and $m\ge15$, the field at its inner edge exceeds its horizon value by $10^{11}$ or more, and thus the relative horizon flux error grows ($7\times10^{-2}$ at $m=15$, $N=48$). These horizon fluxes are below $10^{-23}$ of the corresponding \scri~fluxes, so the total flux is unaffected. 

\bibliography{bibitems}

\end{document}